# LAGRANGIAN COHERENCE AND SOURCE OF WATER OF LOOP CURRENT FRONTAL EDDIES IN THE GULF OF MEXICO

Luna Hiron[1], Philippe Miron[1], Lynn K. Shay[2], William E. Johns[2], Eric P. Chassignet[1], Alexandra Bozec[1].

[1] Center for Ocean-Atmospheric Prediction Studies, Florida State University, Florida, USA.
[2] Rosenstiel School of Marine, Atmospheric, and Earth Science, University of Miami, Florida, USA.
Corresponding author: Luna Hiron (lhiron@fsu.edu)

**ABSTRACT**

Loop Current Frontal Eddies (LCFEs) are known to intensify and assist in the Loop Current (LC) eddy shedding. In addition to interacting with the LC, these eddies also modify the circulation in the eastern Gulf of Mexico by attracting water and passive tracers such as chlorophyll, Mississippi freshwater, and pollutants to the LC-LCFE front. During the 2010 Deepwater Horizon oil spill, part of the oil was entrained not only in the LC-LCFE front but also inside an LCFE, where it remained for weeks. This study assesses the ability of the LCFEs to transport water and passive tracers without exchange with the exterior (i.e., Lagrangian coherence) using altimetry and a high-resolution model. The following open questions are answered: (1) How long can the LCFEs remain Lagrangian coherent at and below the surface? (2) What is the source of water for the formation of LCFEs? (3) Can the formation of Lagrangian coherent LCFEs attract shelf water? Strong frontal eddies leading to LC eddy shedding are investigated using a 1-km resolution model for the Gulf of Mexico and altimetry.

The results show that LCFEs are composed of waters originating from the outer band of the LC front, the region north of the LC, and the western West Florida Shelf and Mississippi/Alabama/Florida shelf, and potentially drive cross-shelf exchange of particles, water properties, and nutrients. At depth ($\approx$ 180 m), most LCFE water comes from the outer band of the LC front in the form of smaller frontal eddies. Once formed, LCFEs can transport water and passive tracers in their interior without exchange with the exterior for weeks: these eddies remained Lagrangian coherent for up to 25 days in the altimetry dataset and 18 days at the surface and 29 days at depth ($\approx$ 180 m) in the simulation. LCFE can remain Lagrangian coherent up to a depth of $\approx$ 560 m. Additional analyses show that the LCFE involved in the Deepwater Horizon oil spill formed from water near the oil rig location, in agreement with previous studies. Temperature-salinity diagrams from a high-resolution model and aircraft expendable profilers show that LCFEs are composed of Gulf of Mexico water as opposed to LC water. Therefore, LCFE formation and propagation actively modify the surrounding circulation and affect the evolution of the flow and the transport of oil and other passive tracers in the Eastern Gulf of Mexico.

1. **Introduction**

The Loop Current (LC) is part of the North Atlantic western boundary current system and contributes to the transport of warm water from the tropics to the subtropics and higher latitudes. The warm and salty LC flows from the Caribbean Sea into the eastern Gulf of Mexico (GoM) through the Yucatan Straits, turns anticyclonically to form a loop, and exits the GoM through the Florida Straits, where it turns into the core of the Florida Current (Fig. 1). Thereby, the LC connects the Caribbean Sea, the GoM, and the North Atlantic, and plays an essential role in the transport and exchange of heat, salt, and marine organisms such as larvae and *Sargassum* between these basins (Tester et al., 1991; Lee and Williams, 1999).

The LC has three main phases: (i) the retracted phase or port-to-port scenario, in which the LC presents the shortest path from the Yucatan to the Florida Straits, (ii) the growing phase from retracted to extended, and (iii) the fully extended state when the LC reaches its maximum intrusion in the GoM. During the last phase, the LC becomes unstable and detaches an anticyclonic eddy, called a Loop Current Eddy (LCE), that translates westward toward Mexico at $\approx$ 2.5-6 cm s$^{-1}$ (Lee and Mellor, 2003; Schmitz, 2005). The LC then returns to the retracted phase, and a new LC extension cycle starts.

The LCE shedding occurs at irregular intervals of 6 to 17 months (Vukovich, 1988; Behringer et al., 1977; Sturges and Leben, 2000), which makes forecasting these events quite challenging (Dukhovskoy et al., 2015; Zeng et al., 2015; Wang et al., 2019; Wang et al., 2021). Nevertheless, frontal eddies have been observed to amplify at the neck of the LC and precede LCE detachments, playing an essential role in the LCE shedding (Cochrane, 1972; Vukovich et al., 1979; Vukovich and Maul, 1985; Vukovich, 1988; Lee et al., 1995; Fratantoni et al., 1998; Zavala-Hidalgo et al., 2003; Schmitz, 2005; Shay et al., 2011; Hiron et al., 2020).

Loop Current Frontal Eddies (LCFEs) are cyclonic, cold-core eddies that travel along the periphery of the LC. Their formation mechanisms and characteristics differ from one side to the other of the LC. LCFEs on the western flank of the LC are predominantly generated through barotropic instability and present smaller horizontal scales and higher-frequency variability, whereas frontal eddies on the eastern flank tend to be larger, have lower-frequency variability, and are predominantly generated through baroclinic instability (Chérubin et al., 2006; Hamilton et al., 2016; Donohue et al 2016a; Donohue et al., 2016b; Garcia-Jove et al., 2016; Sheinbaum et al., 2016; Jouanno et al., 2016; Yang et al., 2020). LCFEs participate in the LCE shedding by propagating westward and squeezing the neck of the LC, leading to an eddy pinch-off. Other processes such as vortex merging, vortex stretching, vortex alignment, and LCFE interactions with the LC front can cause further intensification of these LCFEs (Cochrane 1972; Zavala-Hidalgo et al. 2003; Walker et al. 2011; Le Hénaff et al. 2012; Hiron et al., 2020).

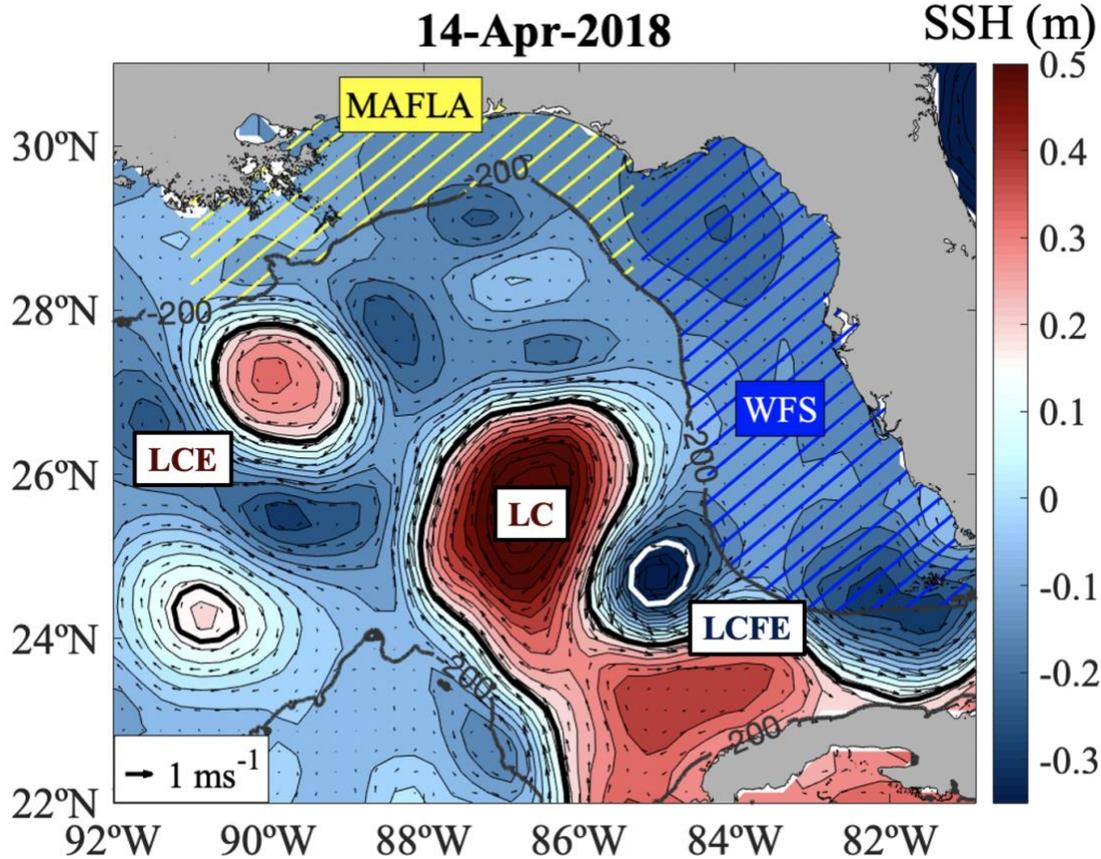

Figure 1. Sea surface height field from altimetry for the Loop Current system, which is composed of the Loop Current (LC), Loop Current Eddy (LCE), and Loop Current Frontal Eddy (LCFE). The LC and LCE are shown by the 17 cm SSH black contour (Leben, 2005) and the LCFE by the -28 cm SSH white contour (Hiron et al., 2020). To better identify the features, the white color in the colormap is centered on 17 cm, which is the boundary of the LC as defined by Leben (2005). This background is shown in all subsequent figures. The 200 m isobath is indicated by the solid gray line, and the Mississippi/Alabama/Florida (MAFLA) shelf and the West Florida Shelf (WFS) are shown by the yellow and blue hatched area, respectively.

Besides playing an important role in the LCE shedding, strong frontal eddies can strengthen the LC front, tilt the isopycnals, increase the horizontal density gradient, and modify the balance of forces in the LC front, shifting from a geostrophic balance to a gradient-wind balance regime associated with the increase in the centrifugal force (Hiron et al., 2020; Hiron et al., 2021).

In addition to modifying the LC, strong frontal eddies can also affect the circulation in the eastern Gulf of Mexico. The LC-LCFE front can attract passive tracers and light particles (Olascoaga and Haller, 2012), such as chlorophyll (Toner et al., 2003; Olascoaga et al., 2013) and Mississippi freshwater (Schiller and Kourafalou, 2014; Androulidakis et al., 2019), and potentially *Sargassum* and larvae, from the shelf to offshore regions. Maps of Finite Size Lyapunov Exponent identified ridges of strong horizontal stretching in the LC-LCFE front and around the LCFEs, suggesting LCFEs do extract mass from the surroundings (Hiron et al., 2020). The attraction of fluid by cyclonic-anticyclonic dipoles, such as an LC-LCFE pair, is well known. However, an LCFE was observed to attract fluid not only to the LC-LCFE front but also into the center of the eddy during the 2010 Deepwater Horizon oil spill (Walker et al., 2011). The frontal eddy intensified on the north flank of the LC and played a vital role in entraining and then carrying oil in its interior for weeks, which prevented the oil from reaching the Florida Keys and polluting the coast and its ecosystems (Fig. 1.2 in Walker et al., 2011; Liu et al., 2011; Olascoaga and Haller, 2012; Gonçalves et al., 2016).

Lagrangian coherence describes the tendency for eddies to carry particles in their interior without exchange with the exterior. Although this LCFE was observed to carry oil in its center (Walker et al., 2011), a complete study to characterize the ability of the LCFEs to transport water and passive tracers from one place to another without exchange with the exterior (e.g., Lagrangian coherence) is still lacking. Lagrangian coherent vortices are efficient in transporting particles, nutrients, water mass properties, and pollutants such as oil. Eulerian methods such as SSH isolines, relative vorticity, Okubo-Weiss parameter, and Rossby number often require a subjective threshold and have been shown to fail to identify the Lagrangian coherent boundary of eddies (Haller, 2005; Haller and Beron-Vera, 2013; Beron-Vera et al., 2013). When advected forward in time, Eulerian contours have been observed to stretch, filament, and leak the material inside the contour to the background flow, losing coherence rapidly (Beron-Vera et al., 2013). To solve this issue, Haller and Beron-Vera (2013) developed a method to identify Lagrangian coherent vortices fully independent from the observer's viewpoint (frame invariant) based on the Cauchy-Green strain tensor field. Their method was applied to Agulhas eddies and LCEs and found that these vortices can remain Lagrangian coherent for months (Haller and Beron-Vera, 2013; Beron-Vera et al., 2018).

The goals of this manuscript are to (1) assess the Lagrangian coherence of the LCFEs at and below the surface based on the method of Haller and Beron-Vera (2013), (2) evaluate the source of water that forms the Lagrangian coherent LCFEs, and (3) investigate, qualitatively, the potential ability of these eddies to drive cross-shelf exchanges, with attraction of shelf water to offshore areas. To achieve these goals, this study uses altimetry and a 1-km resolution, non-assimilative HYCOM simulation for the Gulf of Mexico. The ability to represent part of the submesoscale field is an important aspect of this model. Drifter trajectories and chlorophyll maps were used to support the results found with the model and altimetry, and airborne profilers provided information on the water mass properties of the LCFEs. The focus is on the LCFEs associated with LCE shedding events since these tend to be larger, stronger and more organized than other LCFEs (Hiron et al., 2020).

The structure of the manuscript is as follows: Sections 2 and 3 describe the datasets and methodology, respectively; section 4 assesses the LCFE Lagrangian coherence using the high-resolution HYCOM simulation; section 5 investigates the source of water that forms the LCFEs; section 6 repeats the analyses in sections 4 and 5 using altimetry-derived geostrophic velocities during the 2010 *Deepwater Horizon* oil spill and another case in 2004. In section 7, the potential temperature ($T_\theta$)-salinity (S) properties of the LCFEs are evaluated; in section 8 drifter trajectories and chlorophyll distributions are used to confirm the previous findings; and section 9 contains the concluding remarks.

## 2. Datasets

### 2.1. Altimetry SSH and geostrophic velocities

The analyses conducted in this manuscript use the satellite altimeter reprocessed global ocean gridded L4 sea surface heights and derived geostrophic velocities (SEALEVEL_GLO_PHY_L4_REP_OBSERVATIONS_008_047) processed by the DUACS multi-mission altimeter data processing system and distributed by the Copernicus Marine and Environment Monitoring Service (CMEMS). This product (1/4° resolution) is computed with respect to the 2012 twenty-year-long mean dynamic topography and processes data from all altimeter missions (Jason-3, Sentinel-3A, HY-2A, Saral/AltiKa, Cryosat-2, Jason-2, Jason-1, T/P, ENVISAT, GFO, ERS1/2).

We use the sea surface height above the geoid (i.e., the absolute dynamic topography [ADT]) and subtract from it the daily mean ADT over the Gulf of Mexico to remove the thermal expansion/contraction of the upper ocean associated with the seasonal variability. The mean ADT is computed by averaging for each day the ADT field over the Gulf of Mexico deep waters (> 200 m). The final product after removing the mean ADT is referred to as sea surface height (SSH).

### 2.2. 1/100º resolution, non-assimilative Gulf of Mexico HYCOM simulation

In this study, we use a 10-year (1994–2003), high-resolution ($\approx$ 1 km), non-assimilative HYbrid Coordinate Ocean Model (HYCOM) simulation of the Gulf of Mexico, version 2.3.01, forced by absolute surface hourly wind fields from the Climate Forecast System Reanalysis (CFSR), with turbulent heat fluxes computed following the Kara bulk formulation (Fairall et al., 2003; Kara et al., 2005). The sea surface salinity is relaxed to climatology at the surface with a value equivalent to a piston velocity of 15 meters over 30 days. The relaxation is turned off when the difference between the simulation and the climatology is over 0.5 psu to preserve the river runoff water properties. HYCOM uses a hybrid coordinate system, with pressure coordinates near the surface, in the mixed layer, and in unstratified seas, isopycnal coordinates in the open, stratified ocean, and sigma/terrain-following coordinates in coastal regions (Bleck, 2002; Chassignet et al., 2003). The 1-km resolution Gulf of Mexico simulation has 41 vertical hybrid layers and uses a high-resolution (0.01° x 0.01°) bathymetry (version 2.0; Velissariou, 2014), and the HYCOM 1/12º (GOFS3.1) reanalysis for initial conditions (Jan 1st 1994) as well as daily boundary conditions (available at https://www.hycom.org/dataserver/gofs-3pt1/reanalysis).

### 2.3. Aircraft expendable profilers

During the 2010 *Deepwater Horizon* oil spill, a set of nine research flights on NOAA's WP-3D aircraft were conducted between 8 May and 9 July 2010 over the LC to deploy oceanic profilers and assess the position of the LC relative to the well site (Shay et al., 2011). Expendable probes measuring temperature, salinity, and currents were deployed over the LC system during the shedding of LCE Franklin and provided important information about the internal structures of the LC, the LCE Franklin, and the LCFEs, from the surface down to ≈ 1200 m, with a 2 m vertical resolution. A total of 588 airborne profilers were deployed, including 35 airborne expendable conductivity-temperature-depth profilers (AXCTD). From those AXCTD, only deep profilers (down to 1000 m) and profilers far from the Mississippi Fan were used in this manuscript to avoid the freshwater influence, resulting in a total of 33 AXCTD. More information about this dataset can be found in Shay et al. (2011).

### 2.4. Drifter database

Lilly and Pérez-Brunius (2021) regrouped drifter trajectories from various experiments available in the GoM (≈ 2000 in the eastern GoM) from 1992 through 2020 and interpolated them onto an hourly resolution. The dataset is available at https://doi.org/10.5281/zenodo.3985916.

### 2.5. Chlorophyll (MODIS)

The Moderate Resolution Imaging Spectroradiometer (MODIS) is an instrument that measures chlorophyll aboard the Terra (originally known as EOS AM-1) and Aqua (originally known as EOS PM-1) satellites. The Terra and Aqua MODIS instruments sample the Earth's surface every one to two days. The dataset used in this study is the Level 3, in which an 8h rolling mean was applied to obtain a continuous map of chlorophyll. The dataset is available at https://oceancolor.gsfc.nasa.gov/data/aqua/.

## 3. Methodology

### 3.1. Identifying Lagrangian coherent vortices

The theory behind the identification of Lagrangian coherent vortices was developed by Haller and Beron-Vera (2013) and consists in identifying vortices with material (i.e., Lagrangian) boundaries that resist filamentation in 2D fluid flows. By definition, for non-diffusive tracers, no flux occurs through the boundary of a Lagrangian coherent vortex, such that the water inside the advected vortex remains isolated and preserves its properties. Lagrangian coherent eddies are thus efficient in transporting heat, salt, and other properties, as mixing with the exterior water is minimal. Furthermore, Haller and Beron-Vera (2013)'s method is objective, unlike Eulerian approaches where different vortices are identified depending on the observer's viewpoint (Haller, 2005). Objectivity is the property in which a quantity or principle is conserved under a change of the observer's reference frame. Different frames of reference can be chosen when studying eddies, such as an inertial frame, a frame co-rotating with the Earth's surface, or frames co-rotating with individual vortices. An objective method identifies the same vortex for all frames.

Following Haller and Beron-Vera (2013), Lagrangian coherent eddies were sought as fluid regions enclosed by exceptional material loops that defy the typical exponential stretching occurring in unsteady fluids. Such loops $r(s)$ are limit cycles (i.e., closed trajectories) of the vector field $\eta_\lambda^\pm$ and *uniformly* stretch by some amount $\lambda$, where:

$$r'(s) = \eta_\lambda^\pm(r(s)), \quad \eta_\lambda^\pm = \sqrt{\frac{\lambda_2 - \lambda^2}{\lambda_2 - \lambda_1}} \xi_1 \pm \sqrt{\frac{\lambda^2 - \lambda_1}{\lambda_2 - \lambda_1}} \xi_2.$$

Here $s$ is the curvilinear distance along the trajectories, $r'(s)$ is the tangent to the loop, $\lambda$ satisfies $0 < \lambda_1 < \lambda^2 < \lambda_2$, and $(\lambda_j, \xi_j)$ are the $j^{th}$ eigenvalue-eigenvector pair of the Cauchy-Green strain tensor

$$C_{t_0}^{t_0+T}(x_0) = \nabla F_{t_0}^{t_0+T}(x_0)^\top \nabla F_{t_0}^{t_0+T}(x_0),$$

an objective (i.e., observer-independent) measure of the fluid deformation from the initial time $t_0$ to the final time $t_0 + T$. The flow map $F_{t_0}^t$,

$$F_{t_0}^t(x_0) \coloneqq x(t; t_0, x_0),$$

takes a fluid particle from position $x_0$ at time $t_0$ to position $x$ at a later time $t$, and describes the evolution of the fluid. This follows by solving

$$\dot{x} = v(x, t), \quad x(t_0) = x_0,$$

where $v$ is the fluid velocity. The variables $x, x_0$, and $v$ are 2-dimensional. The boundary of a Lagrangian coherent vortex is defined to be the outermost limit cycle of $\eta_\lambda^\pm$. As noted above, all subsets of Lagrangian coherent boundaries uniformly grow or shrink from $t_0$ to $t_0 + T$ by the same amount $\lambda$. While any $\lambda$ is admissible, the case $\lambda \approx 1$ is particularly relevant to incompressible flows. When the flow is incompressible, the area enclosed by a "$\lambda$-loop" is preserved. In this case, which holds for geostrophic velocities exactly and approximately for HYCOM velocities within layers, the condition $\lambda \approx 1$ implies strongly constrained deformation, i.e., maximal coherence. The vortices identified in this manuscript have values of λ close to the optimal value of 1, where stretching is uniform along the loop. We allow the λ parameter to vary between [0.7, 2.0] to allow for some filamentations. For selected vortices, the longest time scale T of a vortex boundary is presented, i.e., no vortex is identified for larger values of T.

Based on Haller and Beron-Vera (2013)'s method, Karrasch et al. (2015) and Karrasch and Schilling (2020) developed an automated procedure to detect Lagrangian coherent vortices in large datasets. Their code, used in this manuscript to detect LCFE coherent boundaries, can be found at https://github.com/CoherentStructures/CoherentStructures.jl. To assess LCFE Lagrangian coherence, the code is applied to two strong LCFEs during different LCE shedding events using the high-resolution HYCOM simulation and five other cases using altimetry. To find the longest time $T$ over which the eddy remains coherent before the LCE shedding ($T = t_f - t_0$), the final time $t_f$ is set as the time of the shedding and the initial time $t_0$ was reduced iteratively until no loop is identified. The time of the shedding $t$ was determined using the 17 cm SSH contour, which is representative of the boundary of the LC (Leben 2005). The effective diameter $D$ of the Lagrangian coherent eddies was computed from the area $A$ inside the Lagrangian coherent boundary, where $D = 2\sqrt{\frac{A}{\pi}}$.

### 3.2. Advection of particles

The source of water for the formation of the LCFEs was investigated by populating the area inside the Lagrangian coherent boundaries at the final time $t_f$ with approximately six thousand particles and advecting them backward in time for $T + 60$ days ($T + 110$ days for the altimetry dataset), where $T$ is the number of days each LCFE remained Lagrangian coherent. The particles were advected for longer for the altimetry dataset because velocities from altimetry tend to be weaker than in the model, and the particles take longer to get advected. The particle advection was accomplished by applying a 4$^{th}$ order Runge-Kutta (RK) method to the daily velocity field output for both model and altimetry with a 6h time step and linear interpolation. For the model, trajectories were computed within layers. HYCOM has isobaric coordinates near the surface, isopycnal coordinates at depths, and terrain-following coordinates in coastal regions. Particles reaching terrain-following coordinates were removed. This was done by removing any particles varying more than 0.5 kg m$^{-3}$ in the isopycnal layers, and 0.01 m in the isobaric layers.

### 4. Assessment of LCFE Lagrangian coherence from a model perspective

Using the 1 km-resolution HYCOM simulation, the material coherence of two strong LCFEs associated with two LCE shedding events is presented ($t_f$ = 25 June 1999, $t_f$ = 8 February 2001). For the 1999 case, the LCFE remained materially coherent for 18 days at the surface before the complete detachment of the LCE, from 7 June 1999 to 25 June 1999 (Table 1; Fig. 2a). In other words, the coherent portion of the LCFE conserved and carried the water inside its boundary for 18 days without exchange with its surroundings. The coherence is even stronger below the surface. For layers 16 ($\approx$ 100 m) and 23 ($\approx$ 180 m inside the coherent boundary), the eddy remained coherent for 23 and 29 days before the shedding event, respectively (Fig. 2b,c). Layer 1 and 16 are isobaric and represent, respectively, the surface and the 100 m layer. By contrast, layer 23 is isopycnal and the depth varies considerably. The duration of Lagrangian coherence for this eddy increases with depth and reaches a maximum at layer 23. Below that layer, the coherence time starts to decrease until layer 28 ($\approx$ 560 m inside the coherent boundary), with a maximum Lagrangian coherence time of 6 days (Table 1). Below layer 28, no Lagrangian coherent boundary was detected.

For the 2001 case, the LCFE remained coherent for 10 days at the surface before the shedding event, from 29 January 2001 to 8 February 2001, and for 15 days for layer 16 ($\approx$ 100 m; Fig. 2d,e). Differently from the 1999 case, the 2001 case presented the longest coherence (19 days) at layer 24 ($\approx$ 260 m; Fig. 2f). Notice that the LCFE Lagrangian coherent boundaries have time scales around a third to half of the time it takes a developing meander crest and trough to propagate along the north and east sides of an extended LC (40 days – 100 days; Donohue et al., 2016a).

The frontal eddies are coherent for longer periods at depth compared to the surface for both cases. Note that the area enclosed by the loops at deeper layers ($\approx$ 180 m–260 m) is larger than those at the surface, which means that the frontal eddies are able to transport larger quantities of particles and for longer periods at deeper layers ($\approx$ 180 m–260 m) compared to the surface. The decrease in material coherence at the surface compared to the lower layers is likely due to high-frequency motions such as winds, near-inertial motions, Ekman transport, ageostrophic motions, and submesoscale variability that contribute to eddy incoherence (Weisberg et al., 2001; Curcic et al., 2016; Beron-Vera et al., 2019).

The eddy's vertical coherence is assessed for the 1999 case (Fig. 3). The LCFE Lagrangian coherent boundaries were computed from the surface until layer 27 for a fixed Lagrangian coherent time $T$ of 10 days, which was the largest common Lagrangian coherent time among these layers. Fig. 3 shows the volume of water associated with the LCFE, that remained Lagrangian coherent for 10 days, superposed to the isopycnal layers 23 ($\sigma_\theta = 1026.6$ kg m$^{-3}$), 25 ($\sigma_\theta = 1027.0$ kg m$^{-3}$), and 27 ($\sigma_\theta = 1027.3$ kg m$^{-3}$) for 18 June 1999. The isopycnal layers are deeper in the LC and shallower in the LCFEs. The difference in depths in the same layer in the LC versus LCFE can reach values of 500 m, in agreement with the strong horizontal density gradient and isopycnal tilting observed in the LC-LCFE front from a mooring array (Hiron et al., 2020).

The LCFE Lagrangian coherent boundaries become larger with depth until 200 m. Below 200 m, the boundaries decrease in size until 440 m (layer 27) for $T = 10$ days. Without setting a fixed $T$, we found that the LCFE was coherent until layer 28 ($\approx$ 560 m), with a maximum $T$ of 6 days (Table 1). Layer 28 has potential density of 1027.4 kg m$^{-3}$ and a near constant temperature of 6.8°C. Layer 29, which has potential density of 1027.4 kg m$^{-3}$ and temperature of 5.8°C, did not present a Lagrangian coherent boundary near the location of the LCFE. The 6°C isotherm is usually taken as the base of the LC (Bunge et al. 2002; Hamilton et al. 2018). Thus, the LCFE is Lagrangian coherent only until the isopycnal of the base of the LC ($\approx$ 1000 m), which corresponds to an average depth of 560 m inside the LCFE.

The average diameter of the Lagrangian coherent portion of the frontal eddies is 85 km ± 2 km at the surface for the 1999 case. The diameter increases with depth until it reaches a maximum diameter of 104 km ± 3 km for layer 23 ($\approx$ 180 m) and layer 24 ($\approx$ 220 m), then decreases with depth until layer 28 ($\approx$ 560 m) with diameters of 60 km. For the 2001 case, the average diameter is 60 km ± 2 km at the surface and 113 km ± 1 km for layer 24 ($\approx$ 260 m). Overall, the diameters varied between 57 km and 85 km at the surface in the simulation. For reference, previous work based on Eulerian methods (altimetry SSH) have found LCFE diameters at the surface to be ≈ 80–120 km on the northern and eastern flank of the LC (Vukovich and Maul, 1985; Le Hénaff et al., 2014). In general, Lagrangian-based diameter estimates, where vortex boundaries are identified as material loops that stretch uniformly, are smaller than their Eulerian observer-dependent counterparts. Similarly, Lagrangian-based life expectancy estimates are shorter.

**Table 1.** Information about the Lagrangian coherence of two LCFE cases at and below the surface from the numerical simulation, and five cases at the surface from altimetry: Maximum time of Lagrangian coherence, and minimum, maximum, average, and standard deviation of the coherent boundary diameter.

|  |  | LCFE | Layer | Average depth inside LCFE | Maximum time of Lagrangian coherence (days) | Diameter (km) | | | |
|---|---|---|---|---|---|---|---|---|---|
|  |  |  |  |  |  | min | max | mean | std |
| Model |  | 2001 case (8 Feb. 2001) | 1 | surface | 10 | 57 | 64 | 60 | 2 |
|  |  |  | 16 | ≈ 100 m | 15 | 64 | 68 | 67 | 1 |
|  |  |  | 24 | ≈ 260 m | 19 | 112 | 114 | 113 | 1 |
|  |  | 1999 case (25 June 1999) | 1 | surface | 18 | 79 | 88 | 85 | 2 |
|  |  |  | 16 | ≈ 100 m | 23 | 71 | 79 | 76 | 3 |
|  |  |  | 23 | ≈ 180 m | 29 | 100 | 109 | 104 | 3 |
|  |  |  | 24 | ≈ 220 m | 25 | 100 | 107 | 104 | 2 |
|  |  |  | 25 | ≈ 300 m | 23 | 97 | 101 | 99 | 1 |
|  |  |  | 26 | ≈ 320 m | 14 | 49 | 50 | 49 | 1 |
|  |  |  | 27 | ≈ 440 m | 10 | 54 | 56 | 55 | 1 |
|  |  |  | 28 | ≈ 560 m | 6 | 60 | 60 | 60 | 0 |
| Altimetry | 2010 | LC shedding (24 May 2010) |  | surface | 20 | 67 | 84 | 78 | 6 |
|  |  | Oil Spill (29 June 2010) |  | surface | 20 | 49 | 60 | 54 | 3 |
|  |  | Third Eddy (29 June 2010) |  | surface | 20 | 61 | 65 | 63 | 2 |
|  | 2004 | Segment 1 (9 May 2004) |  | surface | 23 | 50 | 72 | 60 | 7 |
|  |  | Segment 2 (3 June 2004) |  | surface | 25 | 53 | 69 | 61 | 5 |

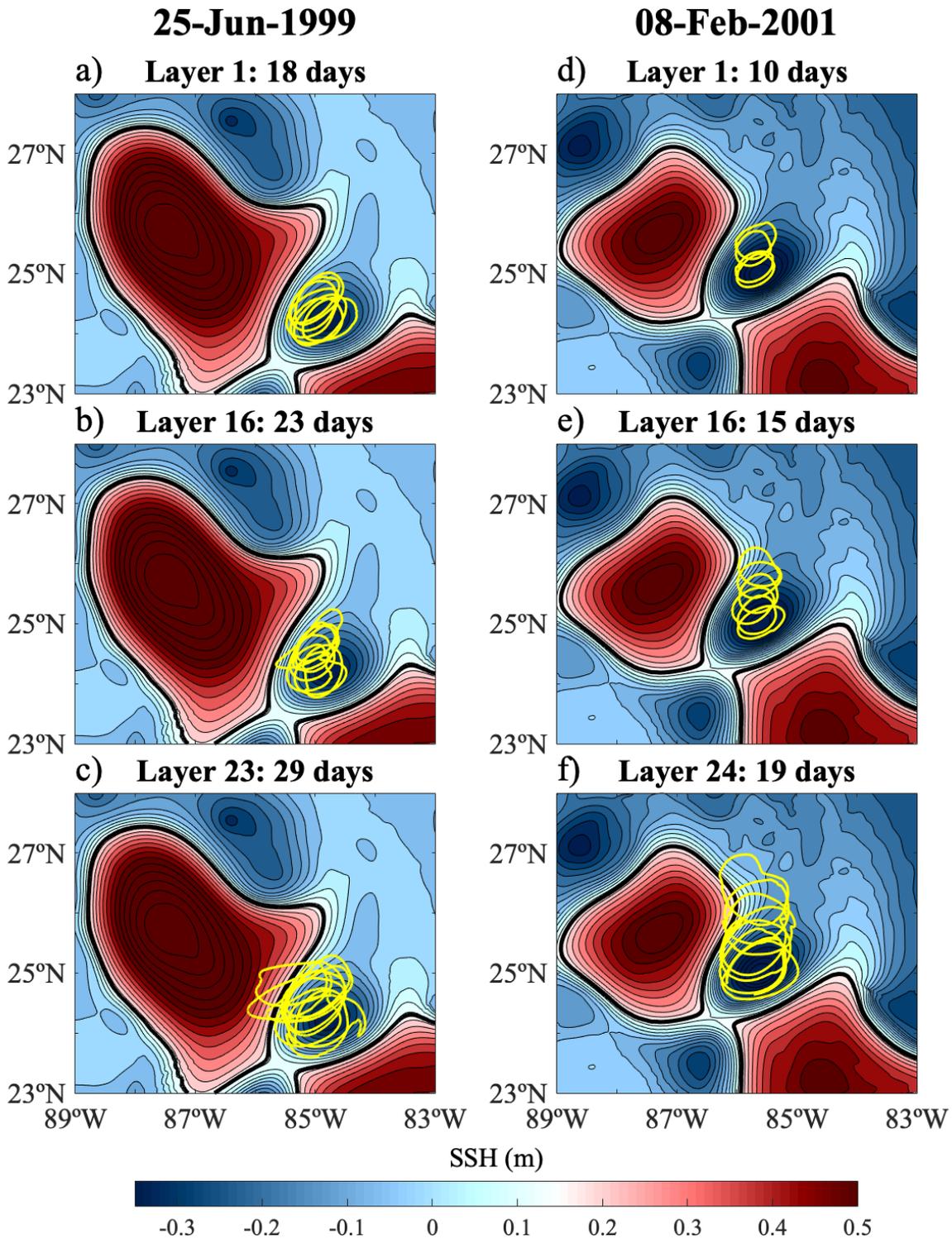

Figure 2. LCFE Lagrangian coherent boundaries (yellow lines) every 3 days using a 1 km-resolution HYCOM simulation for the (a,b,c) 1999 and (d,e,f) 2001 cases, for (a,d) layer 1 at the surface, (b,e,) layer 16 ≈ 100 m, and (c) layer 23 ≈ 180 m and (f) layer 24 ≈ 260 m. Sea surface height maps for the time of the LCE shedding (25 June 1999 and 8 Feb. 2001) are displayed as background for reference. The maximum number of days that the LCFEs remained coherent for each layer is indicated in each subplot.

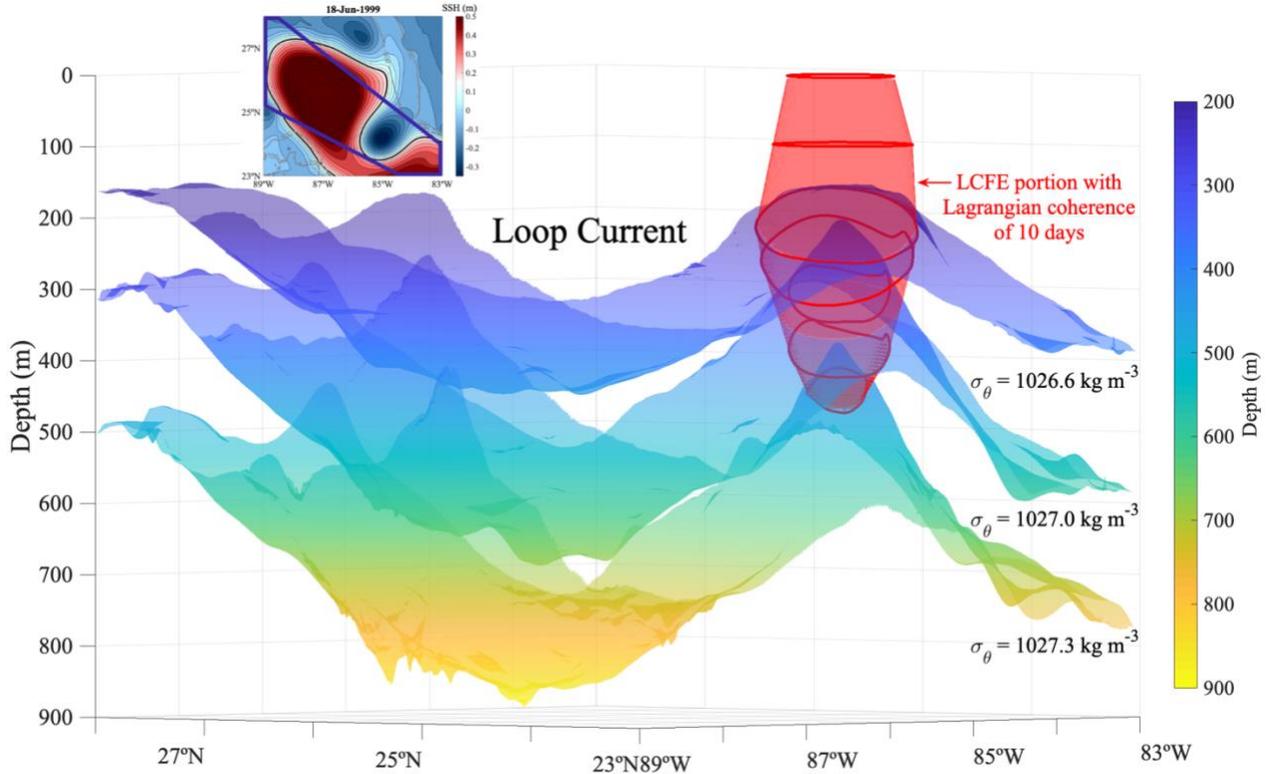

Figure 3. (Red object) Volume enclosing the LCFE Lagrangian coherent boundaries at different layers (red loops) for a time coherence of 10 days superposed to the depth of the isopycnal layers 23 ($\sigma_\theta = 1026.6$ kg m$^{-3}$), 25 ($\sigma_\theta = 1027.0$ kg m$^{-3}$), and 27 ($\sigma_\theta = 1027.3$ kg m$^{-3}$) for 18 June 1999 from the high-resolution HYCOM simulation.

## 5. Source of water of LCFEs from a model perspective

To determine the origin of the water inside the LCFEs, ≈ 6000 particles were initialized inside the LCFE coherent boundary at the final time $t_f$ for both 1999 and 2001 cases and tracked backward in time using a 4$^{th}$ order Runge-Kutta method. The particles are expected to remain together for $T$ days, following the theory, from the final time $t_f = t_0 + T$ to the initial time $t_0$, and then disperse backward into the eastern GoM. For both 1999 and 2001 cases at the surface, the particles remained together for $T$ days, as expected: 18 days for the 1999 case and 10 days for the 2001 (Fig. 4a,b and 5a,b).

The particles are advected further backward in time for a total of 60 days. From $t_0$ to $t_0 - 60$ days, the particle trajectories indicate that the LCFEs build up from the convergence of water from near the LC front, the region north of the LC, and the West-Florida shelf (WFS) and Mississippi/Alabama/Florida (MAFLA) shelf. At time $t_0 - 60$ days, especially for the 1999 case, many particles can be seen on the WFS and MAFLA shelf before crossing the 500 m isobath and converging to form the coherent LCFE (see movie in Supplementary Material). It is important to notice that the shelf particles are not exclusively attracted offshore by the LCFEs but instead by a complex eddy field north of the LC.

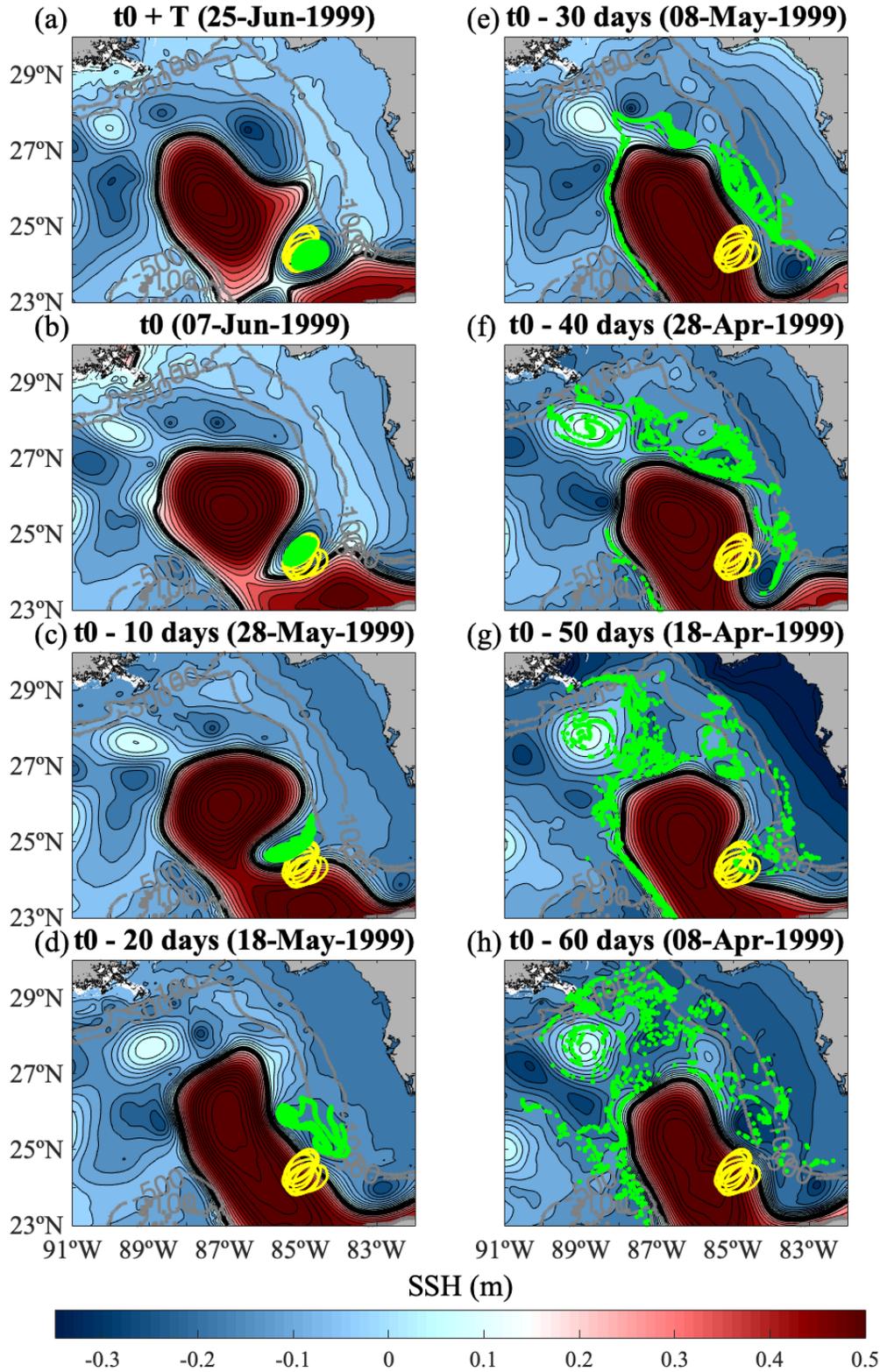

Figure 4. HYCOM sea surface height maps with the location of the particles (green dots) advected backward in time from the final time $t_f = t_0 + T$ to $t_0$ - 60 days for the 1999 LCFE case at the surface. The yellow loops are the Lagrangian coherent boundaries plotted every 3 days.

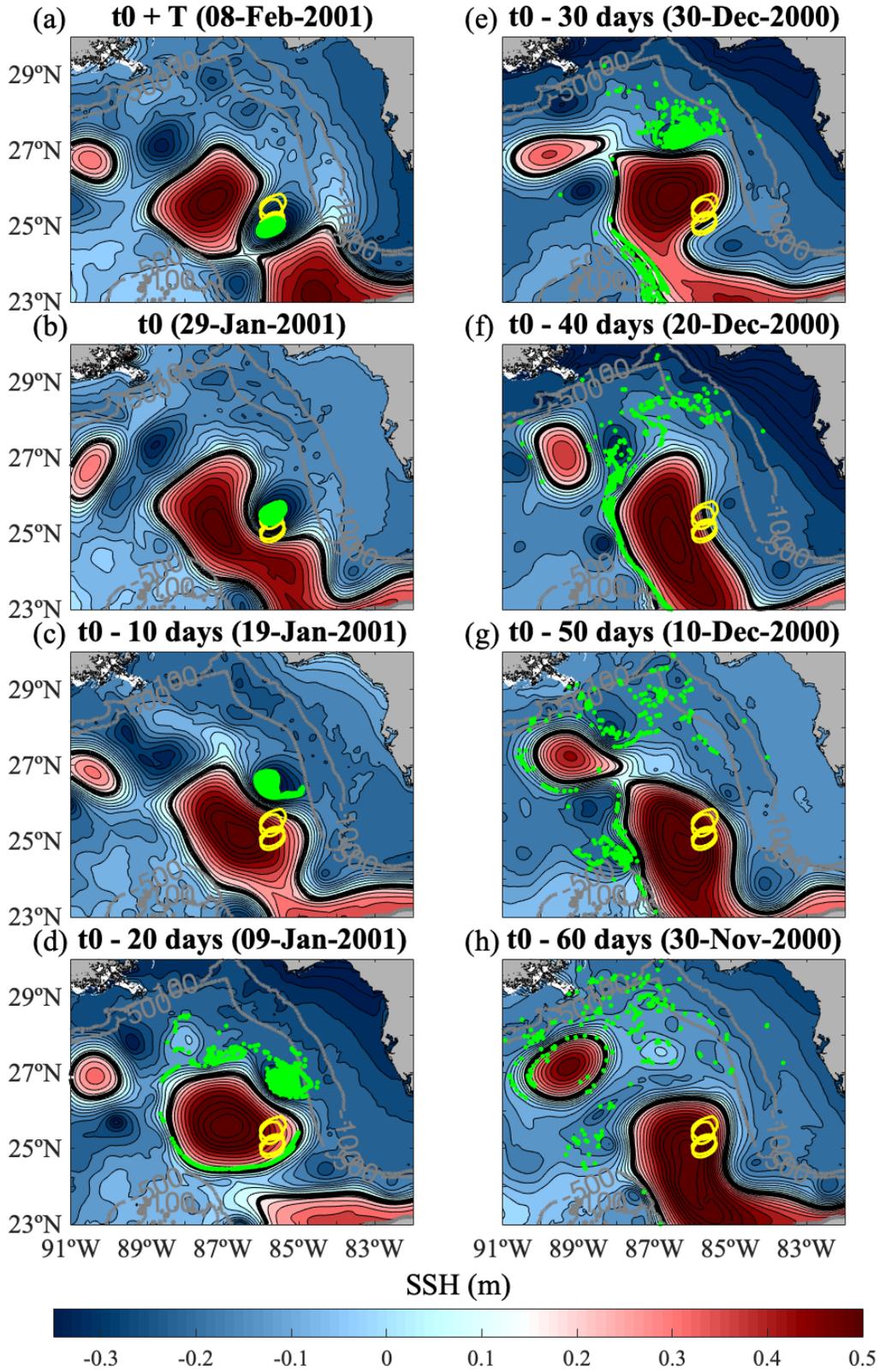

Figure 5. Same as Fig. 4 except for the 2001 LCFE case at the surface. The yellow loops are the coherent boundaries plotted every 3 days.

Further analyses are needed to confirm that LCFEs facilitate cross-shelf exchanges compared to periods without LCFEs. Cross-shelf exchanges are particularly important for bringing water rich in nutrients from the coast to oligotrophic areas offshore. Additionally, fresh water from the Mississippi river can also make its way into the interior GoM (Schiller and Kourafalou, 2014; Androulidakis et al., 2019) and alter the water masses through mixing processes. The attraction of particles from the region north of the LC into the LCFE is in agreement with the observations during the 2010 Deepwater Horizon (Walker et al., 2011).

Although particles are observed to cross the western WFS isobath (Fig. 4), it is important to note that these particles were already close to the 100 m isobath from the start, and no particles originated from the interior of the WFS. In fact, a satellite-tracked drifter study revealed the existence of an isolated, drifter-free region in the southern WFS, denominated the "forbidden zone". This region extends in a triangular-shaped area from South Florida, from the coast to approximately 240 km offshore, to the Tampa Bay area (Yang et al., 1999). The forbidden zone is delimited by a vertically coherent cross-shelf transport barrier that suppresses particles to flow from near the southwest Florida coast to offshore regions (Olascoaga et al., 2006; Olascoaga, 2010). The existence of the forbidden zone could explain the absence of particles coming from near the Florida coast, as opposed to the MAFLA shelf where many particles reach the Mississippi coast (Fig. 4 and 5). Olascoaga et al. (2006) observed a seasonal movement of the transport barrier, with an offshore shift in winter and an onshore shift in summer. The 1999 case (Fig. 4) occurred during mid/end of the spring; thus, the cross-shelf transport barrier was shifted onshore on the WFS. The particles crossing the isobaths possibly come from a north-to-south WFS flow studied in Weisberg et al. (2005). In section 8, a chlorophyll map supports this statement and shows water from the along-WFS flow being entrained inside a frontal eddy located in the neck of the LC.

The same particle analysis was applied below the surface for layer 16 ($\approx 100$ m) and layer 23 ($\approx 180$ m) for the 1999 case (Fig. 6 and Fig. 7, respectively). For both cases, the particles remained together for $T$ days (from $t_0 + T$ to $t_0$), as expected. Unlike the surface, the particles advected (from $t_0$ to $t_0 - 60$ days) in layers 16 and 23 experienced less spreading, which is expected as the surface variability is much higher due to stronger flows and the presence of high-frequency motions such as the winds, near-inertial motions, Ekman transport, ageostrophic motions, and submesoscale variability (Weisberg et al., 2001; Curcic et al., 2016; Beron-Vera et al., 2019). For layer 16, approximately half of the particles that formed the Lagrangian coherent LCFE came from near the LC front, and the other half from the region north of the LC. Since the WFS and the MAFLA shelf are shallower than 100 m, and layer 16 has an averaged depth of 100 m, no particles came from the shelf. For layer 23, a small subset of particles came from the region north of the LC, and most of them come from small frontal eddies that propagate around the LC periphery and eventually merge, especially on the north and northeast flank of the LC, giving rise to a larger, Lagrangian coherent LCFE (see movie in Supplementary Material). Previous studies have observed the merging of frontal eddies in this region (Zavala-Hidalgo et al. 2003; Le Hénaff et al., 2012; Walker et al. 2011).

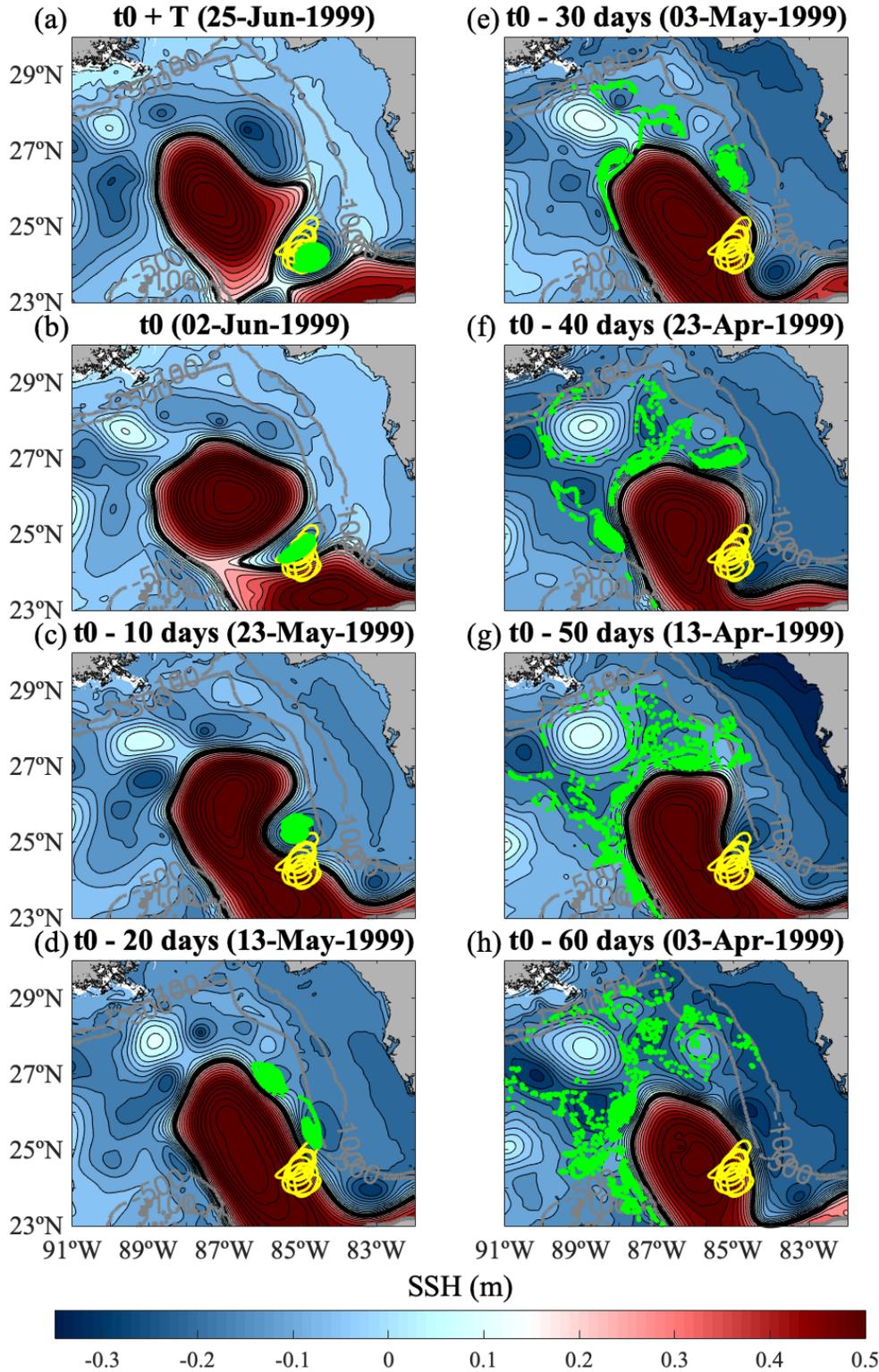

Figure 6. Same as Fig. 4 except for the 1999 LCFE case for layer 16 (≈ 100 m). The yellow loops are the coherent boundaries plotted every 3 days.

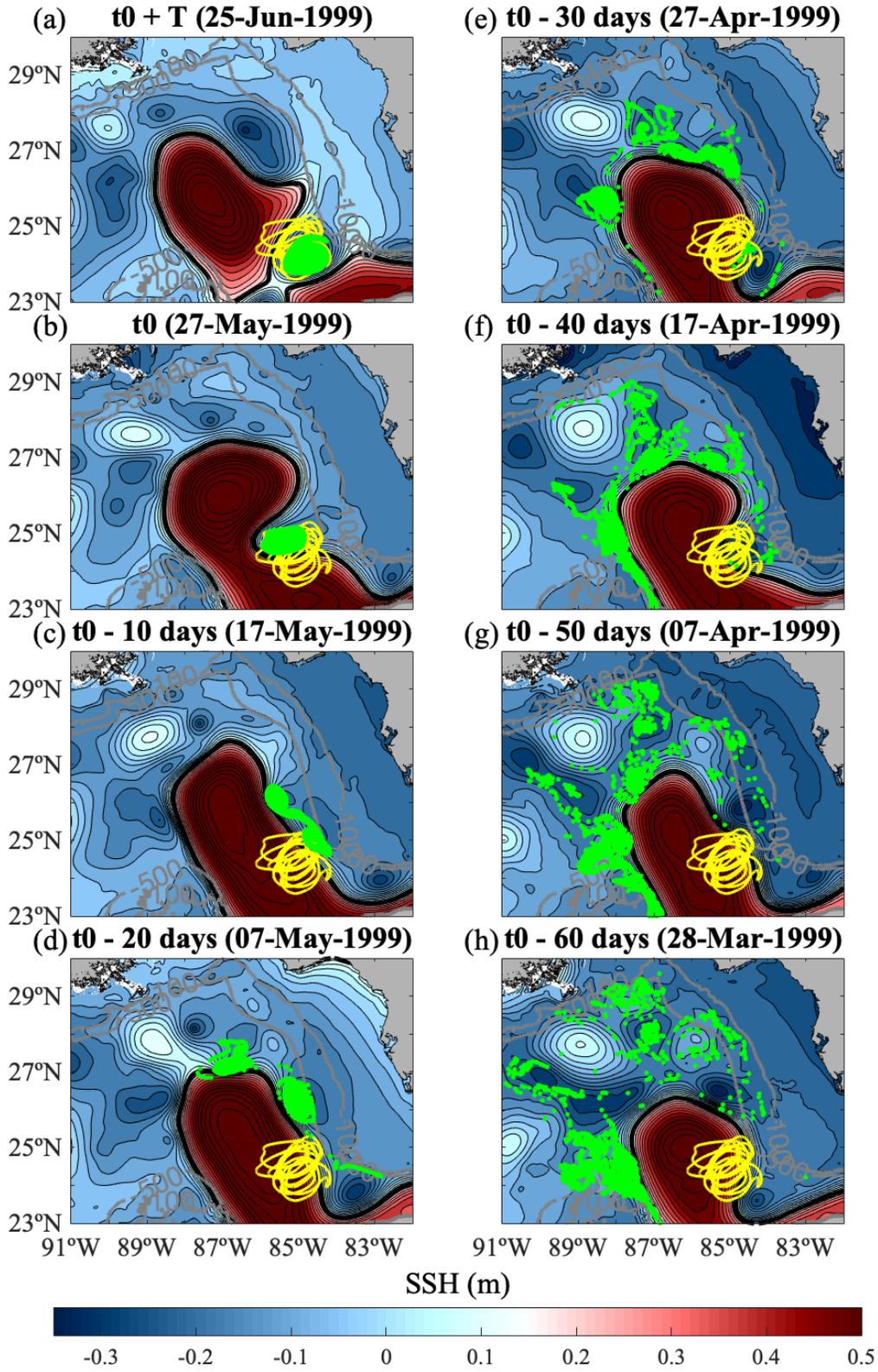

Figure 7. Same as Fig. 4 except for the 1999 LCFE case for layer 23 (≈ 180 m), which is the layer that presented longest Lagrangian coherence (29 days) for this eddy. The yellow loops are the coherent boundaries plotted every 3 days.

It is important to notice that the particle advection scheme is 2-dimensional along HYCOM horizontal coordinates. In isopycnal layers (e.g., layer 23), the vertical motion is accounted for by both the vertical displacement of isopycnal surfaces and the motion along sloping isopycnals. The only component of the vertical velocity neglected is the cross-isopycnal vertical velocity, which we assume to be small. However, in isobaric layers (e.g., layers 1 and 16), the particles are advected at a constant depth without vertical velocities, possibly introducing some error to the particle trajectories in these layers.

From Figs. 4–7, it is unclear whether the particles being advected from near the LC front are part of the LC boundary or located on the outer, cyclonic side of the LC front. Therefore, in section 7, the $T_\theta$-S properties of these particles are further investigated to determine whether the water carried near the LC front is composed of LC water, Gulf water, or a mixture of both. Tracking the particles backward in time from the Lagrangian coherent boundary provided information on the origin of the water that forms the LCFE. To assess the fate of the particles once the eddy loses coherence, e.g., to assess how rapidly the particles escape from the LCFE once its coherent boundary breaks, the particles were integrated forward in time for 60 days starting from time $t_f$ for the 1999 case at the surface (Fig. 8). After time $t_f$, which is when the LCE detaches, the LCFE coherent boundary starts to break down. Most particles from within the former LCFE boundary are entrained on the outer side of the detached LCE, then disperse into the eastern GoM around the LC, and some are entrained on the LC's cyclonic belt and exit the GoM. The LCE was also identified as having a Lagrangian coherent boundary (Beron-Vera et al., 2018), therefore, after its formation, no outside particles get into the LCE core, in agreement with our results.

Layer 23 is shallower inside the LCFE coherent boundary, with an averaged depth of 180 m, and can reach depths of 300 m to 350 m when outside the frontal eddy. The difference in depth between the particles inside and outside the frontal eddy for layer 23 agrees with the theory of vortices in geostrophic/gradient-wind balance where deeper isopycnal layers tend to rise inside cyclonic eddies compared to the surrounding.

The depth of deeper isopycnals layers was observed to also vary within the frontal eddy. Isopycnals tend to be shallower in the portion of the frontal eddy closer to the LC, and deeper in the portion opposite to the LC. Consequently, simulated particles inside the frontal eddies vary in depth from shallower to deeper water in a wavy pattern as the particles rotate within the frontal eddy, with up to 100 m of depth difference (Fig. 5.12 in Hiron, 2021). The same wavy pattern is observed for the kinetic energy along the trajectories (not shown). Particles inside the LCFE presented higher KE in the portion of the frontal eddy closer to the LC and lower KE in the portion opposite to the LC. The asymmetry in KE is characteristic of frontal eddies in general in which the vortex is intensified where in contact with the strong current.

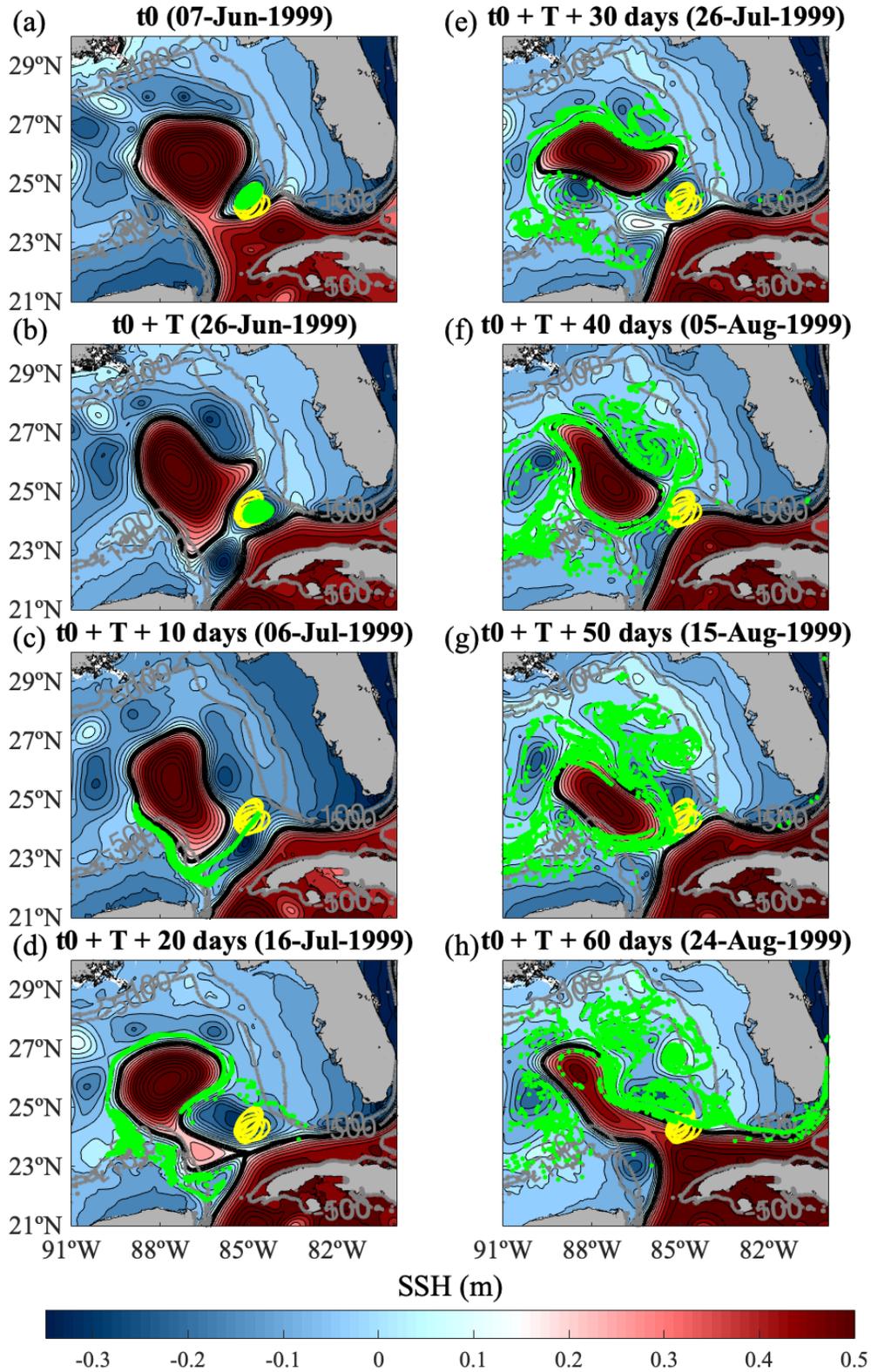

Figure 8. HYCOM sea surface height maps with the location of the particles (green dots) advected forward in time from the initial time $t_0$ to $t_0$ + T + 60 days for the 1999 LCFE case for the surface (layer 1). The yellow loops are the coherent boundaries plotted every 3 days.

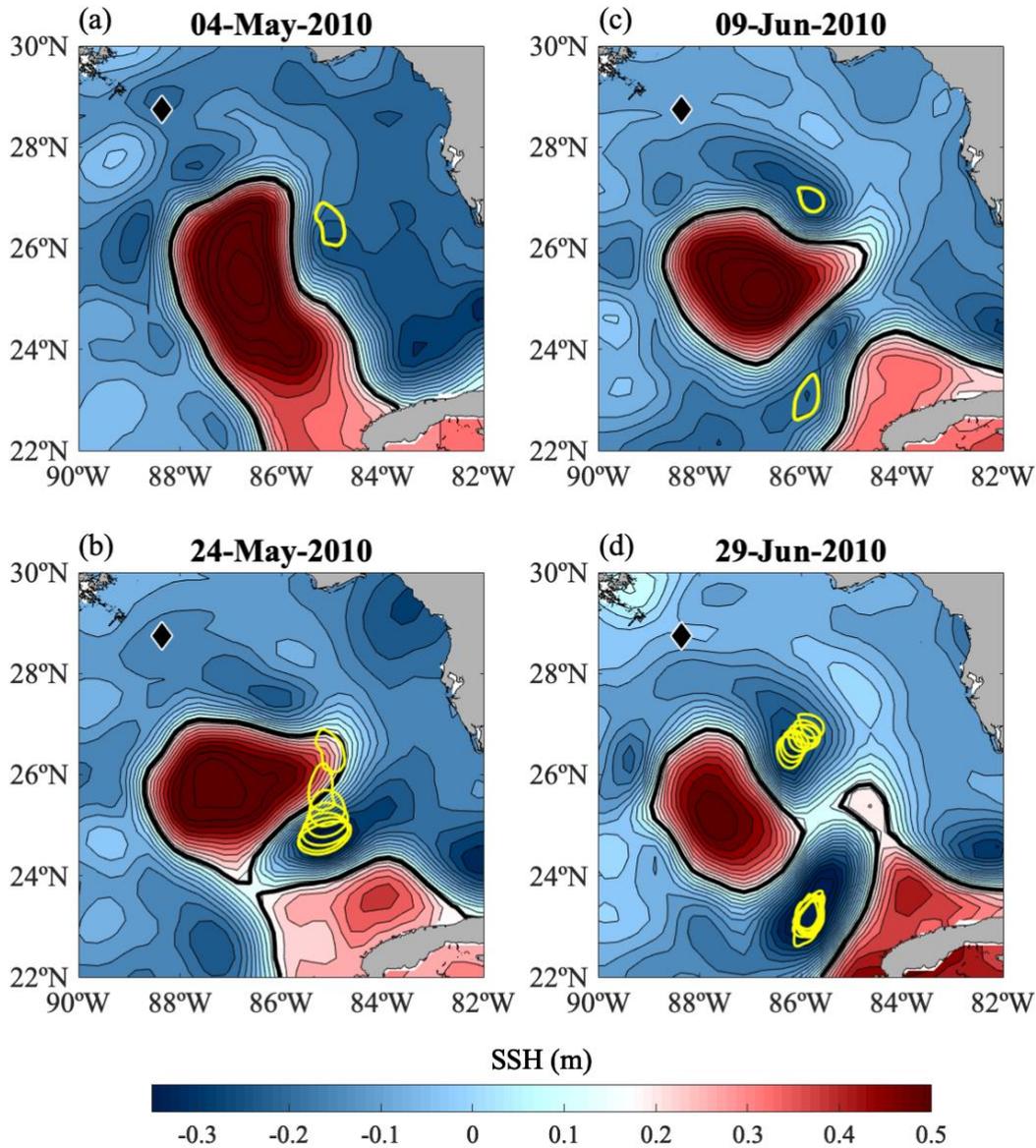

Figure 9. Sea surface height from altimetry for the initial time (a,c) and (b,d) final time of the Lagrangian coherence for three LCFEs detected using altimetry: one LCFE leading to the shedding of eddy Franklin (a,b), and two others leading to eddy Franklin's reshedding (c,d). The yellow line shows the LCFE coherent boundaries at the initial time (a,c) and every three days (b,d). The black diamond indicates the location of the Deepwater Horizon oil rig.

## 6. Assessment of Lagrangian coherence and source of water of LCFEs from an altimetry perspective: the 2010 Deepwater Horizon oil spill and a 2004 case

The Deepwater Horizon rig exploded on April 22, 2010 in the northeast Gulf of Mexico causing oil to leak from the wellhead in the Desoto Canyon, located approximately 41 miles off the coast of Louisiana. The wind and the waves carried part of the oil to the beaches and coast of Louisiana, Mississippi, Alabama, and northwest Florida (Özgökmen et al., 2016). The other part of the oil was carried offshore and entrained in the LC-LCFE front and inside an LCFE along the north flank of the LC (Walker et al., 2011; Olascoaga and Haller, 2012). The oil remained inside the LCFE for the following weeks, preventing it from reaching the Florida Keys and polluting the coast and its ecosystems (Fig. 1.2 in Walker et al., 2011; Liu et al., 2011; Olascoaga and Haller, 2012; Gonçalves et al., 2016).

The Lagrangian coherent vortex analysis was replicated using geostrophic velocities from altimetry for three cases of strong LCFEs during the 2010 Deepwater Horizon oil spill. The first LCFE was the one leading the shedding of LCE Franklin, on 24 May 2010. A month after the detachment, eddy Franklin re-attached to the LC before two other LCFEs, one on each side of the LC's neck, converged toward each other and constricted the LC, leading to a reshedding event. The LCFE located on the northeastern flank of the LCE during the reshedding is the one that trapped a considerable amount of the Deepwater Horizon oil in its interior (Fig. 9).

Coincidently, the three LCFEs — the one leading to the shedding (Fig. 9a,b) and the two others leading to the reshedding (Fig. 9c,d) — remained Lagrangian coherent for 20 days before the shedding. These results were obtained by setting the final time as the shedding time and varying the initial time until the eddy is not coherent anymore, as performed with the model output. The maximum time of Lagrangian coherence for the LCFEs using altimetry (20 days) is larger than the 18 days and 10 days obtained for the 1999 and 2001 cases from the HYCOM simulation at the surface. This difference is to be expected as the altimetry field is smoothed and does not resolve all the surface variability, whereas the 1-km resolution model resolves the high-frequency dynamics associated with winds and inertial waves that can affect eddy coherence. The LCFE along the northeastern flank of the LCE in Fig. 9c,d is responsible for attracting and transporting the Deepwater Horizon oil in its center. The entrainment started a couple of weeks before the eddy boundary became Lagrangian coherent. The location of the oil spill is indicated by the black diamond in Fig. 9. The Lagrangian coherent vortex analysis found that this LCFE remained coherent from 9 June 2010 to 29 June 2010.

In Zavala-Hidalgo et al. (2003), negative SSH anomalies on the west flank of the LC were observed to persist between 1.3 and 9 months in a Hovmöller diagram. In their study, 'LCFEs' were defined as any negative SSH anomaly propagating northward on the west flank of the LC. The discrepancy between their findings and ours comes from the fact that, although LCFEs have negative values of SSH, not all negative SSH anomalies are organized, circular, Lagrangian coherent eddies. Furthermore, negative SSH anomalies do not guarantee that only one eddy is being tracked; multiple small swirls could be propagating together and merging with each other. In fact, the authors noted that the tracked cyclones merged with other smaller ones, and the most long-lasting LCFEs merged with at least three other smaller eddies.

Le Hénaff et al. (2014) analyzed a drifter trajectory that looped cyclonically along the LC's periphery in 2004 for over 7 weeks. This drifter was entrained around a frontal eddy in the north flank of the LC, then propagated around the eddy southward on the east flank of the LC until an almost-LCE shedding event. Although the LC's neck became very narrow with the strengthening of the LCFE, a complete shedding did not occur. Applying the same methodology to this specific case using altimetry geostrophic velocities, we found that this eddy did not remain Lagrangian coherent for the entirety of the 7 weeks (Fig. 10). The eddy was Lagrangian coherent from 16 April 2004 until 9 May 2004, for a total of 23 days. Then, it started to be coherent again from 11 May 2004 until the almost-shedding event on 5 June 2004, for 25 days. Although the eddy remained Lagrangian coherent for a period of time after this near-shedding event, the final time was set as the day the LC's neck was the narrowest, to be consistent with sections 4 and 6. During the 2-day window without coherence, the LCFE appeared to break and then reorganize itself; in the SSH maps from altimetry, the LCFE SSH signature becomes weak, the SSH contour changes from a circular to an ellipse shape, and two SSH maxima can be observed inside the SSH contours.

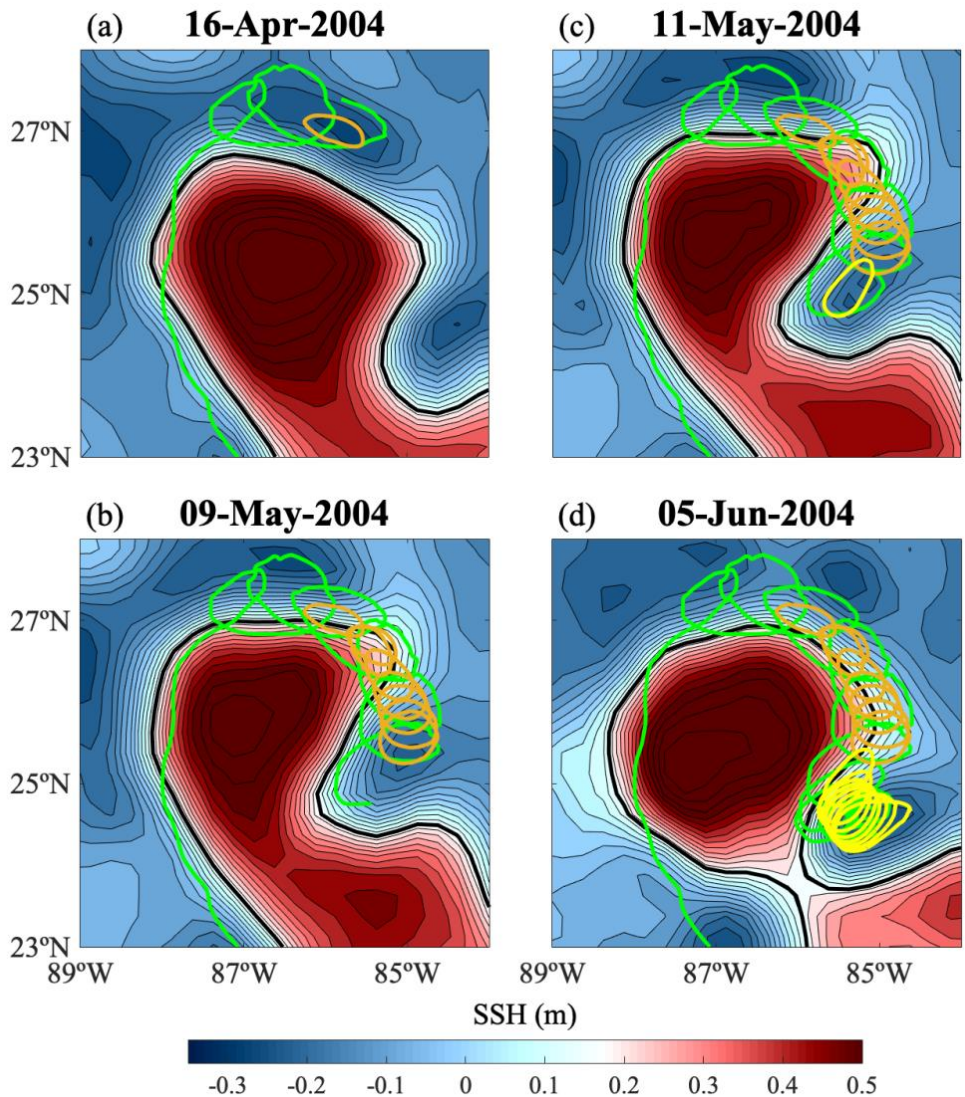

Figure 10. Sea surface height from altimetry for the initial (a,c) and final time (b,d) for two Lagrangian coherent LCFEs in 2004. The green line is the trajectory of the drifter that was entrained around the eddies.

It is important to note that the drifter is looping outside the LCFE's coherent boundary and on the LC's cyclonic belt, where rotational circulation is present. Around 27 May 2004, the drifter was entrained into the LCFE coherent boundary. By definition, no flux occurs through the boundary of a Lagrangian coherent vortex, however, drifter trajectories are also impacted by local winds and waves, which differ from water parcels. Also, elliptic Lagrangian coherent structures are fluid boundaries, whereas a drifter is a physical object that is also influence by its inertia which has been shown to influence considerably their trajectories (Olascoaga et al., 2020).

The diameter of the Lagrangian coherent frontal eddies varied between 49 km and 84 km at the surface in the altimetry dataset (Table 1). Similar to section 4, these values are smaller than the LCFE diameters obtained from Eulerian methods (Vukovich and Maul, 1985; Le Hénaff et al., 2014). The same analysis conducted in section 5 with the backward advection of particles using HYCOM is performed with altimetry for the LCFE leading to the shedding of eddy Franklin (Fig. 9a,b) and the LCFE on the northeastern flank of the LC that attracted the oil to its interior (Fig. 9c,d). The frontal eddy coherent boundary at the final time $t_f$ was populated with particles that were advected backward in time for $T + 110$ days (Fig. 11 and Fig 12). The particles remained inside the Lagrangian coherent boundaries from the final time $t_f = t_0 + T$ to the initial time $t_0$, as expected, then spread (Fig. 11 and Fig 12). For both LCFEs, the particles that build up the frontal eddy coherence originate from the surrounding Gulf water, particularly from the region north and northwest of the LC. When advected long enough ($t_0 - 110$ days), the particles reached the MAFLA shelf (Fig. 11h and Fig 12h), in agreement with the findings using HYCOM. Finally, Fig. 12e shows a line of particles on 25 April 2010 — five days after the rig first broke and started leaking oil to the surface — emanating from the coherent LCFE and passing over the location where the oil spill occurred. This confirms that the altimetry geostrophic field was able to adequately resolve the transport of oil in the offshore region, from around the rig to the interior of the LCFE (Walker et al., 2011; Olascoaga and Haller, 2012).

## 7. Potential temperature ($T_\theta$)-salinity (S) properties of the LCFEs

The $T_\theta$-S properties of the frontal eddies were investigated to identify the source of water for the LCFEs and their water mass signature. First, a $T_\theta$-S analysis was conducted using 33 AXCTD profilers deployed by the NOAA WP-3D research aircraft during the 2010 Deepwater Horizon oil spill, from 8 May to 9 July 2010 (Shay et al., 2011).

The most distinctive water mass in the LC water, in comparison to the Gulf water, is the North Atlantic Subtropical Underwater (NASUW), which is formed by subduction in the North Atlantic subtropical gyre and has a high salinity core (> 36.6 psu) located around the 25.5 kg m$^{-3}$ isopycnal and potential temperature of ≈ 22 °C (Elliott, 1982; Merrell and Morrison, 1981; Morrison et al., 1983; Vidal et al., 1992, 1994; Portela et al. 2018; Hamilton et al., 2018). The Gulf water, on the other hand, has a characteristic subsurface water mass, the Gulf Common Water (GCW), that presents similar temperature but lower salinity values (< 36.5 psu; Elliott, 1982; Vidal et al., 1992, 1994; Portela et al. 2018; Hamilton et al., 2018) than the NASUW. Thus, we use the GCW and NASUW salinity and temperature cores to distinguish between the Gulf and the LC water.

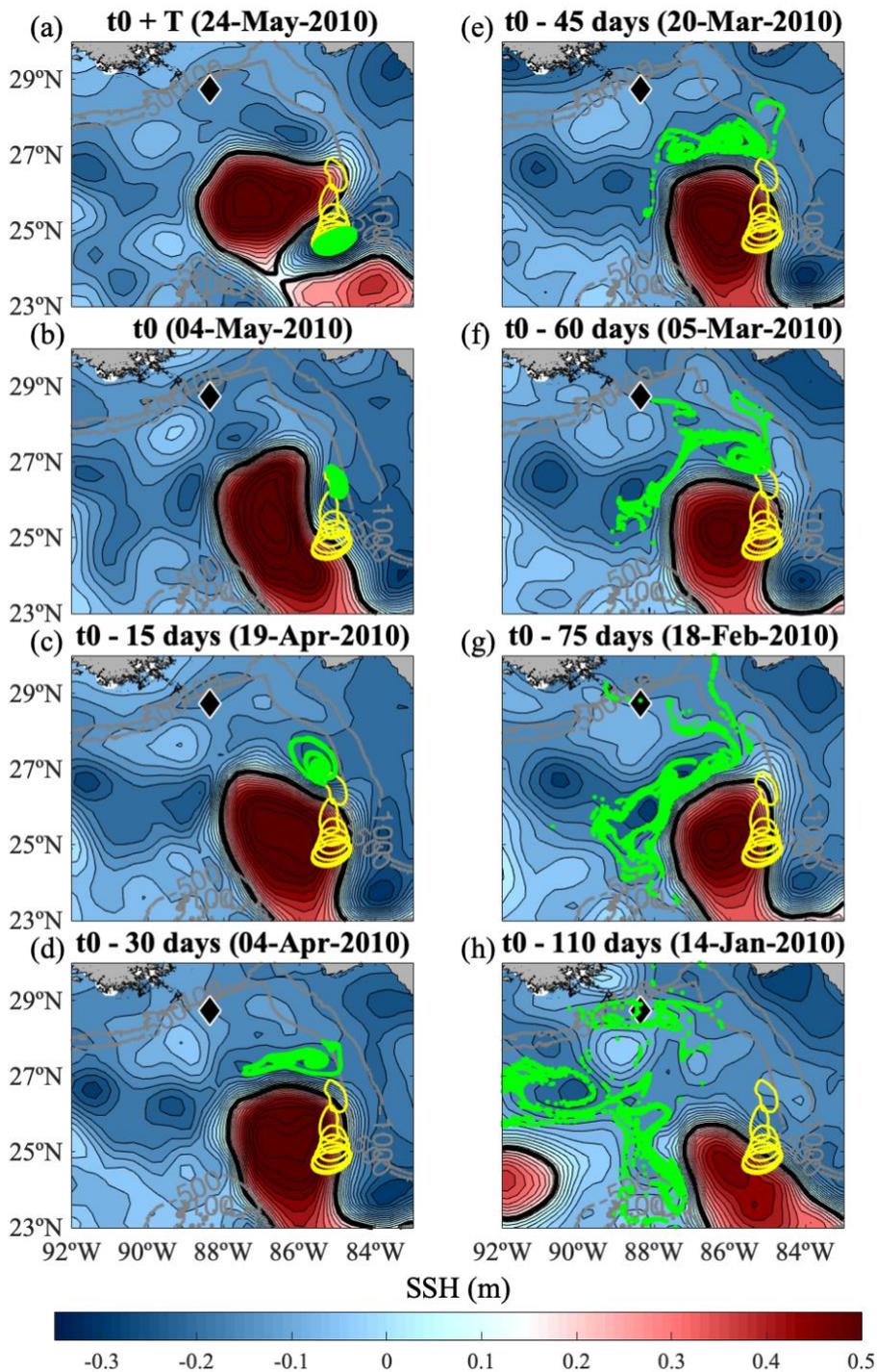

Figure 11. Sea surface height from altimetry with the location of the particles (green dots) advected backward in time using altimetry geostrophic velocities from the final time $t_f = t_0 + T$ to $t_0$- 110 days for the LCFE leading to the shedding of LCE Franklin in 2010. The yellow loops are the LCFE Lagrangian coherent boundaries plotted every 3 days, and the black diamond is the location of the Deepwater Horizon oil spill that exploded on 20 April 2010. The 100 m and 500 m isobaths are indicated by the solid gray lines.

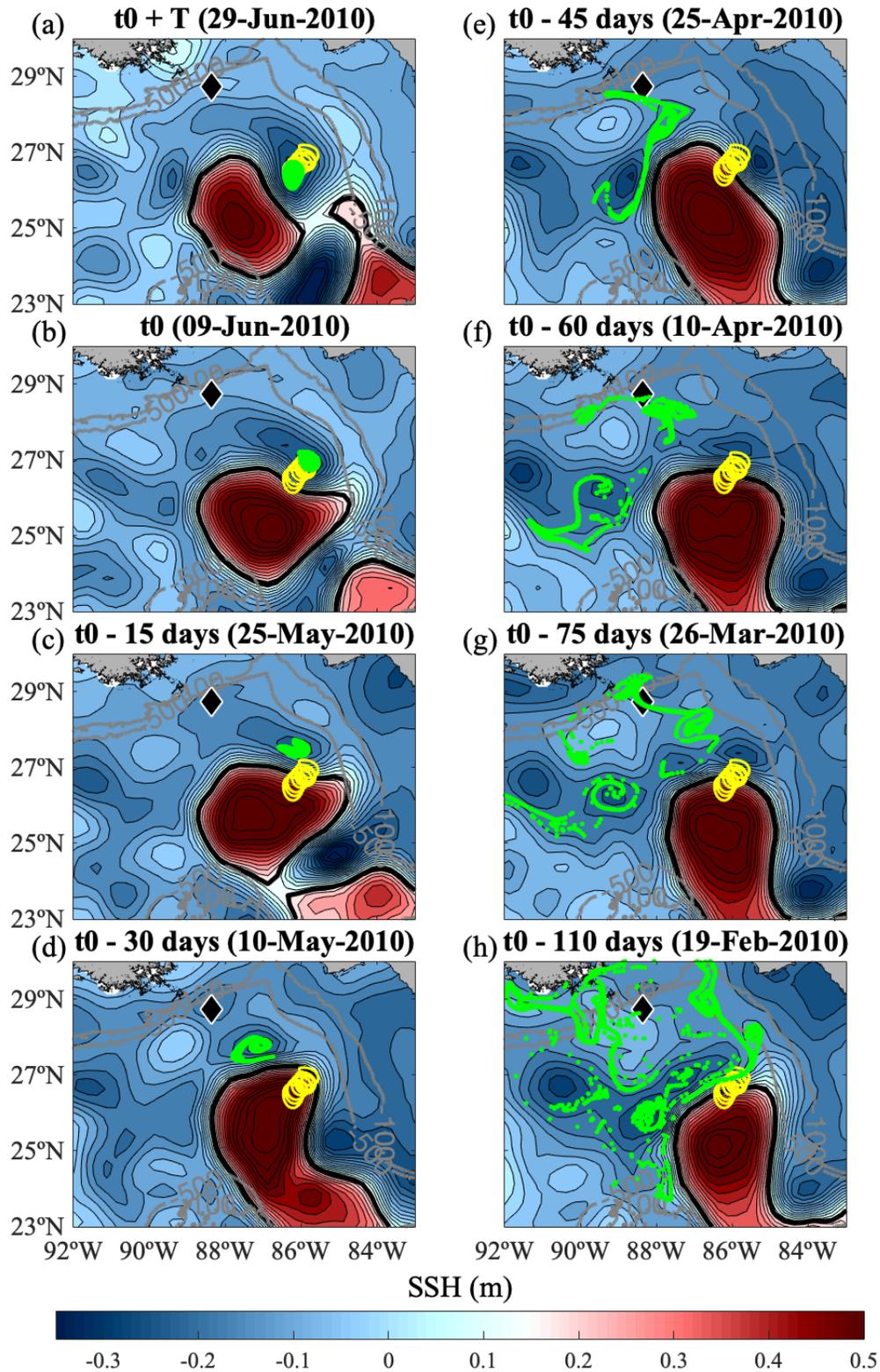

Figure 12. Same as 11, but for the LCFE transporting oil during the 2010 Deepwater Horizon oil spill.

The AXCTD profiles were divided into four environments — LC, LC Front, LCFEs, and Gulf water — based on the location of deployment relative to SSH fields (see Fig. 10 in Shay et al., 2011). The LC environment also includes profilers in the LCE Franklin. The cores of the Gulf Common Water (GCW), the North Atlantic Subtropical underwater (NASUW), and the Sub-Antarctic Intermediate water (SAAIW) (Jaimes and Shay, 2015; and Jaimes et al., 2016) are marked for reference in Fig. 13.

All LC profiles present very homogeneous $T_\theta$-S properties similar to the NASUW core (Fig. 13a), which is expected for the LC and is in agreement with previous literature (Nowlin, 1972; Shay et al., 1998; Jaimes and Shay, 2015; Jaimes et al., 2016). Small $T_\theta$-S variability can be observed at the surface. The Gulf water environment profilers present slightly more variability at depth than the LC but were close to the $T_\theta$-S properties of the GCW core (Fig. 13b). The profiles in the LC front have higher variability, ranging from Gulf water to LC water (Fig. 13c), driven by the high horizontal and vertical shear in this region (Hiron et al., 2020), which causes the mixing of the NASUW and the GCW. Finally, only two profilers sampled an LCFE, and both presented a $T_\theta$-S signature characteristic of the GCW core, similar to the Gulf water profilers (Fig. 13d). Though the sampling of LCFEs was limited, this suggests that the LCFEs are mainly composed of Gulf water, in agreement with previous findings using glider (Rudnick et al., 2015) and expendables (Meyers et al., 2016). The later showed that LCFEs have the same temperature signal as Gulf water but are more stratified. By geostrophic adjustment, the isopycnals experience an upward motion in the center of a strong cyclonic eddy, increasing the stratification. All the profilers sampled SAAIW at depth.

The same $T_\theta$-S analysis is done using the high-resolution HYCOM simulation for selected portions of the LC system for 7 June 1999, which corresponds to the first day of the 18-day Lagrangian coherence for the LCFE (see green circles on the map in Fig. 14a). The LCFE is sampled inside the Lagrangian coherent boundary. Approximately 3000 profilers from the model simulation sampled each environment from the surface down to 1000 m. For all the environments, the simulated profiler salinity values in the upper layer, near the GCW core, are consistently lower by $\approx$ 0.15 psu compared to in-situ measurements (Fig. 13) and previous studies assessment (Nowlin, 1972; Jaimes and Shay, 2015; Jaimes et al., 2016). Models usually have an offset salinity due to uncertainties associated with atmospheric forcing products, the use of climatology at the surface, and the temperature and salinity properties at the boundaries. However, as we are interested in the dynamics and the water properties of the LCFEs compared to the other environments, thus, the salinity offset is not critical for this analysis.

The LC profiles from the model are very homogenous as well, and the temperature variability is similar to the LC expendable profilers. Due to the lower values in salinity, the LC water properties in the model do not overlap with the NASUW core but are closer to the NASUW core than the GCW core. The Gulf water environment presents a wide range of salinity, likely due to the presence of Mississippi freshwater outflow at the surface. Once again, the $T_\theta$-S diagram does not pass through the GCW core but is close. The LC front environment presents a range for $T_\theta$-S values, varying from the LC to the Gulf water environments, as seen with the expendable profilers. Finally, the LCFE $T_\theta$-S properties are very similar to the Gulf water environment and GCW core, in agreement with the findings based on the expendable profilers.

In the upper ocean, LC and LCFE are composed of NASUW and GCW, respectively, which are distinguishable by their salinity. However, below the 26.5 kg m$^{-3}$ sigma-theta ($\sigma_\theta$), the water mass in the LC and LCFE is the same (Fig. 13 and Fig. 14) and consists of Tropical Atlantic Central Water (TACW; Cervantes-Díaz et al., 2022). Although LC and LCFE share the same water mass, the TACW is found at deeper layers in the LC compared to the LCFE. For example, the 26.5 kg m$^{-3}$ layer is at ≈ 400 m in the LC, ≈ 235 m in the LC front, ≈ 165 m in the Gulf water environment, and ≈ 120 m in the LCFE (Fig. 13). By geostrophic (or cyclogeostrophic) adjustment, the isopycnals move downwards in anticyclonic features (LC) and upwards in cyclonic features (LCFE). Thus, due to the vertical displacement of the water mass, the horizontal density gradient is maintained in the LC-LCFE front even at deep waters ($\sigma_\theta > 26.5$ kg m$^{-3}$). These findings agree with Hiron et al. (2020) in which mooring data shows tilted isotherms in the LC-LCFE front from the surface down to 1000 m.

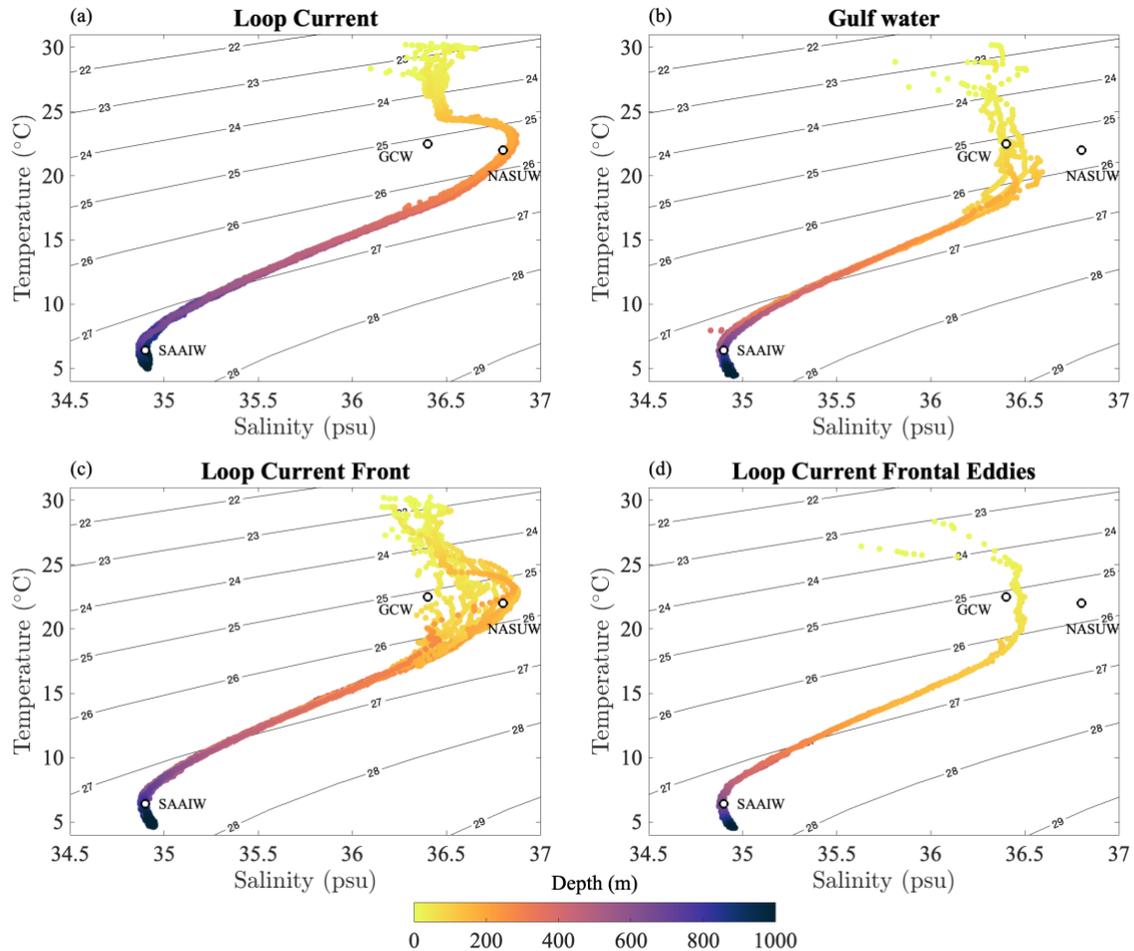

Figure 13. Potential temperature ($T_\theta$)-salinity (S) diagrams using expendable probes from the 2010 oil spill oceanographic mission (8 May to 9 July 2010) for the (a) Loop Current, (b) Gulf water, (c) Loop Current Front, and (d) Loop Current Frontal Eddies. The core of the Gulf Common water (GCW), North Atlantic Subtropical Underwater (NASUW), and Sub-Antarctic Intermediate water (SAAIW) are marked for reference, and the color indicates the depth of the samples. The black lines are contours of constant $\sigma_\theta$.

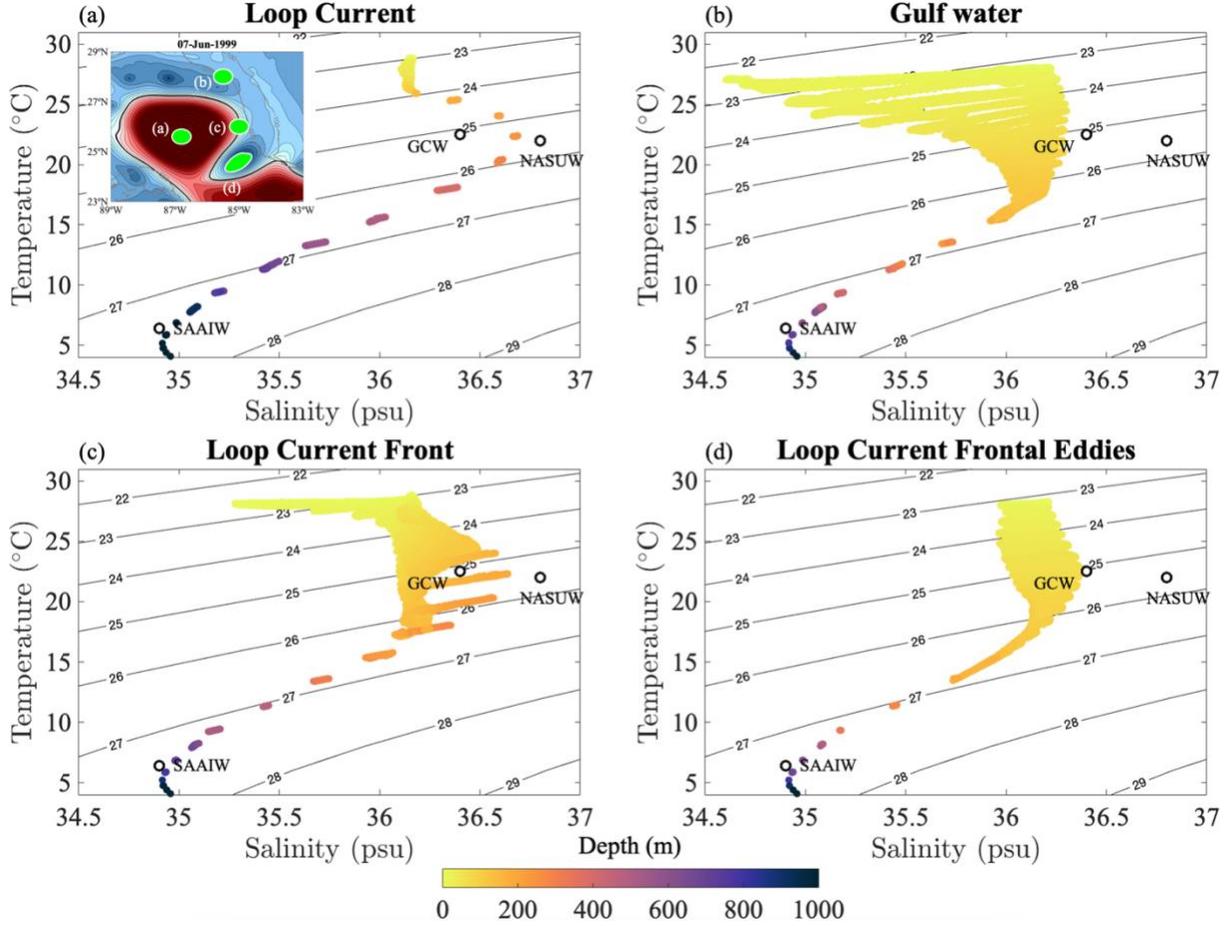

Figure 14. Potential temperature ($T_\theta$)-salinity (S) diagram using the high-resolution HYCOM for the (a) Loop Current, (b) Gulf water, (c) Loop Current Front, and (d) Loop Current Frontal Eddies for 7 June 1999. The core of the Gulf Common water (GCW), North Atlantic Subtropical Underwater (NASUW), and Sub-Antarctic Intermediate water (SAAIW) are marked for reference, and the color indicates the depth of the samples. The black lines are contours of constant $\sigma_\theta$. The locations where the environments were "sampled" in the model are shown on the map on (a) with the corresponding subfigure labels.

     The previous findings (Fig. 13 and 14) provided an overview of the LCFE $T_\theta$-S properties in an Eulerian framework. As a complement, the $T_\theta$-S properties of the particles forming the 1999 LCFE Lagrangian coherence were investigated. The surface layer has strong $T_\theta$-S variability, whereas layer 23 ($\approx$ 180 m) has homogeneous and undistinguishable $T_\theta$-S properties for both the LC and Gulf water profilers. Thus, layer 16 (Fig. 6) appears to be the optimal layer to investigate whether the particles that form the LCFE are composed of GCW (Gulf of Mexico water) or NASUW (LC water). The distribution and density of the particles' $T_\theta$-S properties from 3 April to 25 June 1999 are shown in Fig. 15. The water converging into the LCFE is mostly composed of Gulf water. Nevertheless, it is important to notice that some particles have $T_\theta$-S properties similar

to the LC front environment, with a mix between GCW and NASUW. This again points to strong mixing between GCW and NASUW within the LC front associated with the large horizontal and vertical shears of the flow (Hiron et al., 2020).

A $T_\theta$-S analysis was carried out to find out whether the particles coming from near the LC front (Fig. 6g) are on the cyclonic, outer band of the LC, i.e., strictly composed of Gulf water, or if some are on the anticyclonic side of the front, with LC water properties. For that, a $T_\theta$-S diagram of the particles composing the 1999 LCFE in layer 16 is displayed for a specific time (16 April 1999), when many particles are situated near the LC front (Fig. 16). Notably, all the particles that get into the LCFE come from outside the 17 cm SSH contour, which delimits the boundary of the LC. Additionally, the $T_\theta$-S properties are much closer to the GCW core than the NASUW core, even considering the 0.15 psu shift of the model salinity. Thus, LCFEs are composed of Gulf water coming from the region north of the LC, the WFS and MAFLA shelf, and the outer band, cyclonic side of the LC front.

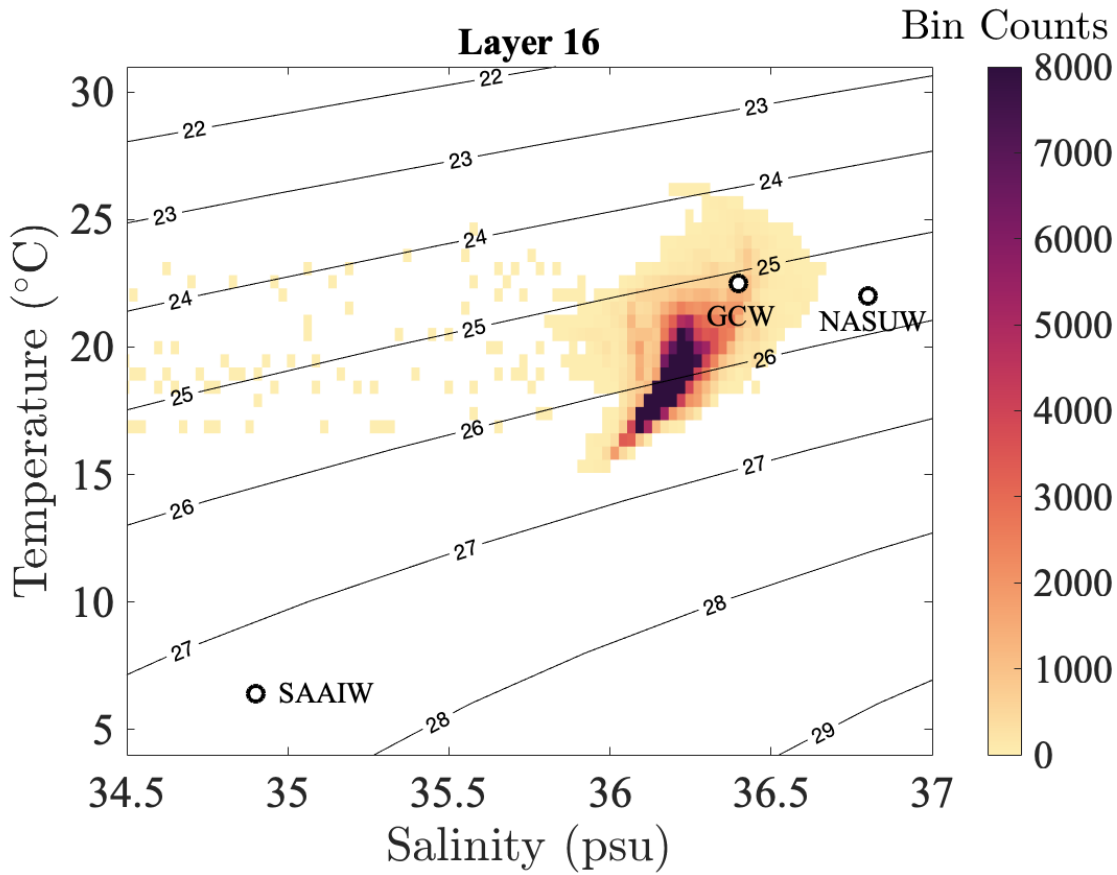

Figure 15. Potential temperature ($T_\theta$)-salinity (S) diagram of the particles forming the 1999 LCFE Lagrangian coherence for layer 16 for the period of 3 April – 25 June 1999 using HYCOM (Fig. 6). The color indicates the density of the particles.

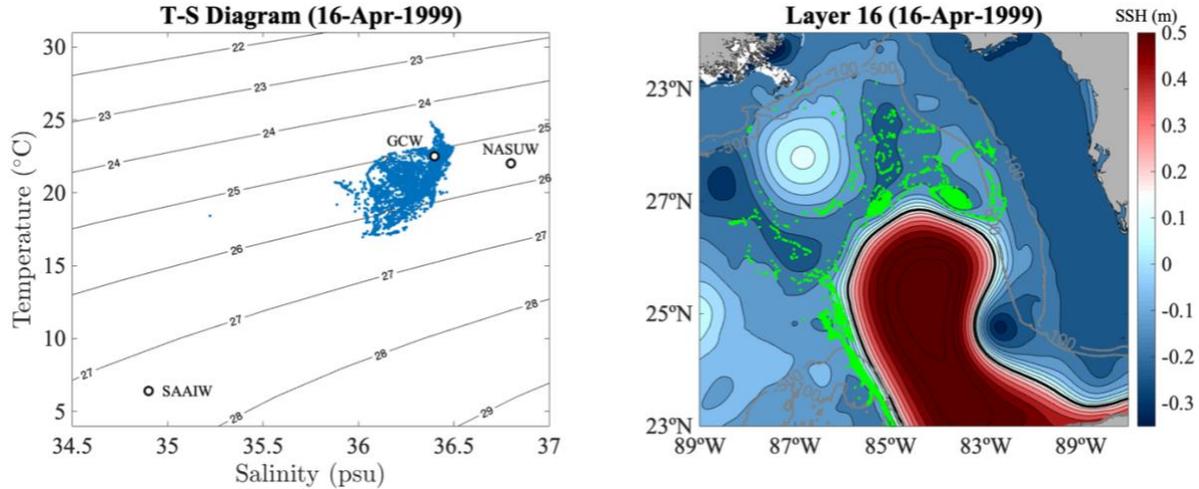

Figure 16. (Left) Potential temperature ($T_\theta$)-salinity (S) for the particles forming the LCFE Lagrangian coherence for the 1999 case for Layer 16, using HYCOM, for a specific time: 16 April 1999. (right) Location of the particles on 16 April 1999.

## 8. Entrainment of surrounding water into LCFEs from observations: drifters and a chlorophyll map

Drifter trajectories and a map of chlorophyll suggests the existence of cross-shelf exchange associated with LCFEs, as seen in sections 5 and 6 using HYCOM and altimetry, respectively. Many drifter trajectories start from the western WFS and MAFLA shelf and are advected offshore around and into the LCFEs (Fig. 17). For the drifters coming from the western WFS and crossing the isobaths (Fig. 17a,b), none of them originated from the inner part of the southwest Florida shelf. This agrees with the "forbidden zone" findings from Olascoaga et al. (2006), in which cross-front transport of particles is suppressed in the southeastern WFS.

The advection of water from the WFS and MAFLA shelf to offshore regions around the LCFE can be better seen in a chlorophyll map (Fig. 18). Chlorophyll behaves approximately as a passive tracer and indicates that coastal water, rich in nutrients, is being exported from the MAFLA shelf and the along-WFS flow to oligotrophic, offshore waters around and into LCFEs. Fig. 18 confirms that the particles crossing the WFS isobaths in Fig. 4 come from a north-to-south along-shelf flow, previously studied in Weisberg et al. (2005), and recently observed in the mean sea surface current product of Lilly and Pérez-Brunius (2021).

Thus, frontal eddies are composed of Gulf water converging from the LC front outer band, the region north of the LC, and the western WFS and MAFLA shelves, and could facilitate cross-shelf exchanges of particles, water properties, and nutrients. Further analyses are needed to confirm that LCFEs drive cross-shelf exchanges.

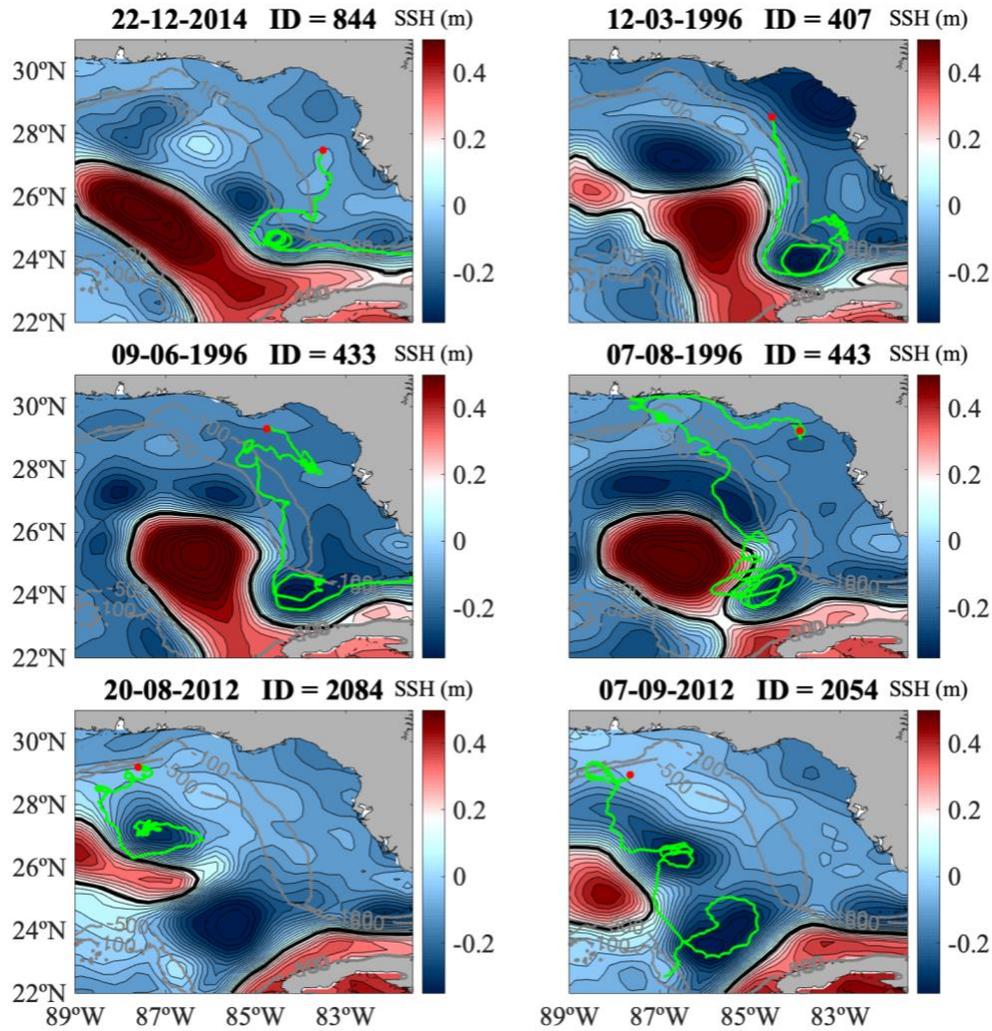

Figure 17. Series of drifter trajectories superposed on sea surface height maps from altimetry showing the entrainment of drifters from the shelf into offshore LCFEs.

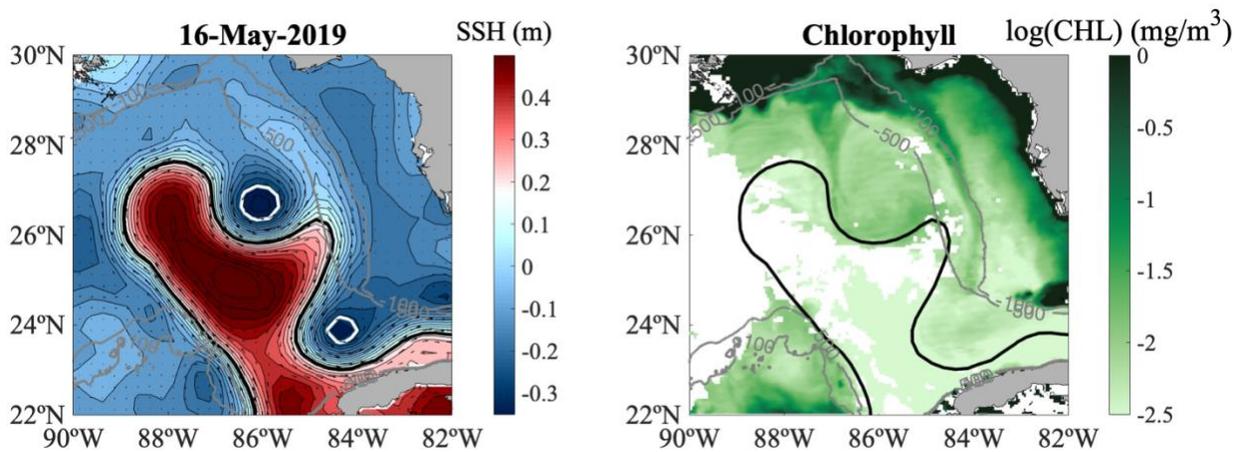

Figure 18. (left) Sea surface height field from altimetry of the Loop Current system on 16 May 2019. (right) Chlorophyll map for 16 May 2019 from the MODIS. Gray lines indicate the 100 m and 500 m isobaths.

## 9. Concluding remarks

This manuscript assessed the Lagrangian coherence of the LCFEs at and below the surface, the source of water that forms these frontal eddies, and their potential ability to attract shelf water offshore in the eastern Gulf of Mexico. Two cases of strong frontal eddies are investigated using a 1-km resolution model for the Gulf of Mexico and 5 cases using altimetry. Only LCFEs associated with the shedding of the LCEs are considered since these eddies tend to be larger and stronger than other LCFEs.

The model analysis for the two cases of strong LCFE leading to LCE shedding showed that frontal eddies can remain Lagrangian coherent for up to 18 days at the surface and up to 29 days at $\approx$ 180 m. Using altimetry, the surface Lagrangian coherence of three LCFEs persisted for 20 days until the detachment of LCE Franklin in 2010, and 23 and 25 days for two other cases in 2004. The year 2010 was particularly important due to the Deepwater Horizon oil spill event and the influence of an LCFE on oil transport. At the surface, the Lagrangian coherent LCFEs presented diameters varying between 49 km and 84 km in the altimetry dataset, and between 57 km and 88 km in the high-resolution simulation. The boundary diameter increases with depth until it reaches a maximum at depth ($\approx$ 180–260 m), with values between 100 km and 114 km, then decreases with depth until layer 28 ($\approx$ 560 m) with diameters of 60 km. The LCFE is Lagrangian coherent only until the isopycnal of the base of the LC ($\approx$ 1000 m), which corresponds to an average depth of 560 m inside the LCFE.

To identify the source of water for the formation of the LCFEs, approximately six thousand particles were placed inside the Lagrangian coherent boundaries and advected backward in time for 60 days for the model simulation and 100 days for the altimetry dataset. The results from altimetry and the model indicate that, at the surface, the LCFEs are composed of waters originating from the outer side of the LC front, the region north of the LC, and the western WFS and MAFLA shelf. At depth ($\approx$ 180 m), in the high-resolution model, the LCFE water comes from the region north of the LC and the outer side of the LC, mostly in the form of smaller frontal eddies that eventually merge on the north flank of the LC and give rise to a larger frontal eddy (see movie in Supplementary Material). $T_\theta$-S analyses using the high-resolution model and aircraft expendable profilers confirm that LCFEs are composed of Gulf water – the water mass characteristic of the Gulf of Mexico – and not of North Atlantic Subtropical Underwater – Caribbean and LC water mass. Drifter trajectories and maps of chlorophyll suggests the existence of cross-shelf exchange associated with LCFEs. Further analyses are needed to confirm and quantify the role of the LCFEs in cross-shelf exchanges.

In summary, the formation of LCFEs actively modifies the surrounding circulation by attracting flow from the outer band of the LC front, the region north of the LC, and the WFS and MAFLA shelf, and could play a role in the cross-shelf exchange of particles, water properties, and nutrients. Once formed, LCFEs can transport water in their interior without exchange with the exterior for weeks. This is particularly important in the case of an oil spill, as oil can remain trapped inside a LCFE for weeks, preventing major environmental disasters. The presence or absence of these Lagrangian features can affect the transport pathways on the periphery of the LC differently. Therefore, LCFE formation and propagation are essential for predicting the evolution of the flow and the transport of oil and other passive tracers in the Eastern Gulf of Mexico.

**Acknowledgments**

L. Hiron would like to acknowledge Francisco Javier Beron-Vera for the discussions about the Lagrangian coherent vortex method that was the base for this manuscript and the comments on the methodology section, and Maria Josefina Olascoaga for the comments on the "forbidden zone". Thank you to Benjamin Jaimes de la Cruz and Peter Hamilton for providing feedback on the water mass analyses and Jonathan Lilly for the interesting discussions on the transport of water by the LCFEs. L. Hiron would also like to acknowledge Matthieu Le Hénaff for suggesting the analysis for the 2004 case and comparison with the drifter trajectory, and three anonymous reviewers that contributed to the improvement of this manuscript. This study has been conducted using E.U. Copernicus Marine Service Information (https://marine.copernicus.eu). This research is made possible through the Gulf of Mexico Research Initiative (GoMRI) through Grant V-487 made to the University of Miami. L. Hiron and L. K. Shay were also supported by the National Science Foundation (NSF) Geosciences under Grant 1941498.

**Supplementary material**

Video #1 (A1_hycom_1999_layer_1):

HYCOM sea surface height maps with the location of the particles (green dots) advected backward-in-time from the final time $t_f = t_0 + T$ to $t_0$- 60 days (and displayed forward-in-time) for the 1999 LCFE case at the surface. The yellow loops are the Lagrangian coherent boundaries plotted every 3 days.

Video #2 (A2_hycom_1999_layer_23):

HYCOM sea surface height maps with the location of the particles (green dots) advected backward-in-time from the final time $t_f = t_0 + T$ to $t_0$- 60 days (and displayed forward-in-time) for the 1999 LCFE case for layer 23 ($\approx$ 180 m inside the LCFE). The yellow loops are the Lagrangian coherent boundaries plotted every 3 days.